\documentclass[10pt]{article}
\usepackage[nofonts]{dgleich-common}

\usepackage[T1]{fontenc}
\RequirePackage[scaled=0.8]{helvet}%
\RequirePackage[scaled=0.75]{beramono}%
\def\lstbasicfont{\fontfamily{fvm}\selectfont}
\makeatletter  
\setboolean{@dgleich@sanssmallcaps}{true}
\makeatother
\usepackage{dgleich-math}
\usepackage{dgleich-tensor}
\newcommand{\ERfull}{Erd\H{o}s-R\'{e}nyi\xspace}

\graphicspath{{./images/}}
\usenatbib
\usehyperref

\makeatletter
\newenvironment{smalleq}
               {\list{}{\leftmargin=0pt \rightmargin=0pt 
               \partopsep=0pt \parsep=0pt \topsep=0pt \itemsep=0pt}%
                \item\relax\footnotesize\abovedisplayskip=0pt\abovedisplayshortskip=0pt}
               {\endlist}
\makeatother     

\begin{document}

\title{The HyperKron Graph Model for higher-order features}

%\author{\IEEEauthorblockN{Nicole Eikmeier and David F. Gleich}
%	\IEEEauthorblockA{ Purdue University\\
%		West Lafayette, USA\\
%		Email: \{eikmeier, dgleich\}@purdue.edu}
%	\and
%	\IEEEauthorblockN{Arjun S. Ramani}
%	\IEEEauthorblockA{Stanford University\\
%		Stanford, USA\\
%		Email: aramani3@stanford.edu}
%}

\author{Nicole Eikmeier, Purdue University \\ Arjun S. Ramani, Stanford University \\ David F. Gleich, Purdue University}
%\address{$^1$Purdue University}
%\author{Arjun S. Ramani}
%\address{$^2$Stanford University}
%\author{David F. Gleich}

\maketitle
\begin{abstract}
Graph models have long been used in lieu of real data which can be expensive and hard to come by. A common class of models constructs a matrix of probabilities, and samples an adjacency matrix by flipping a weighted coin for each entry. Examples include the \ERfull model, Chung-Lu model, and the Kronecker model. Here we present the HyperKron Graph model: an extension of the Kronecker Model, but with a distribution over hyperedges. We prove that we can efficiently generate graphs from this model in order proportional to the number of edges times a small log-factor, and find that in practice the runtime is linear with respect to the number of edges. We illustrate a number of useful features of the HyperKron model including non-trivial clustering and highly skewed degree distributions. Finally, we fit the HyperKron model to real-world networks, and demonstrate the model's flexibility with a complex application of the HyperKron model to networks with coherent feed-forward loops. 
\end{abstract}
	
%\begin{CCSXML}
%	<ccs2012>
%	<concept>
%	<concept_id>10002950.10003624.10003633.10003637</concept_id>
%	<concept_desc>Mathematics of computing~Hypergraphs</concept_desc>
%	<concept_significance>500</concept_significance>
%	</concept>
%	</ccs2012>
%\end{CCSXML}
%
%\ccsdesc[500]{Mathematics of computing~Hypergraphs}
	
%\begin{IEEEkeywords}
%	kronecker graph, graph model, higher-order features, hyperedges
%\end{IEEEkeywords}

\marginnote[26\baselineskip]{Research supported in part by
NSF CAREER CCF-1149756, IIS-1422918, IIS-1546488, NSF Center for Science of Information STC, CCF-0939370, DOE  DE-SC0014543, DARPA SIMPLEX, and the
Sloan Foundation.
}

\section{Introduction}

One of the long-running challenges with network analysis is that there remains a gap between the features of real-world network data and the types of network data that are produced by most efficient synthetic network generators. For instance, simple models such as configuration model~\cite{Bender-1978-configuration} and Chung-Lu~\cite{AielloChungLu2000} are designed to capture the degree distribution of a network, but typically fail to capture any higher-order pattern such as a clustering coefficient. 
%(We explain more about these models in one of the final sections on related work, Section~\ref{sec:related}.) 
Conversely, models that are designed to capture arbitrary features including clustering, such as exponential random graph models, often have exponential computational complexities due to the difficulty of the sampling procedure~\cite{Bhamidi-2011-exponential-mixing}. See for instance a recent survey on the difficulty in generating samples of graphs with the same degree distribution~\cite{Fosdick-configuring}. More structural models, such as stochastic block model, are often designed to test extremely specific hypotheses involving communities and may not be appropriate as more general models. Pragmatic models such as BTER explicitly place clustering in a carefully designed pattern~\cite{Kolda-2014-BTER} at the cost of a larger description of the network.  As an area of active work, this gap between network models and real network data has implications for both studies on the performance of network algorithms when synthetic graphs are used as benchmarks---is it relevant if an algorithm scales well on an unrealistic model of networks?---as well as in the space of hypothesis testing on networks where the synthetic graphs are used as null-models---is it relevant if a feature of a network is a low-probability event with respect to an unrealistic model?

Recently, there has been a surge of interest in higher-order network analysis~\cite{Grilli-2017-higher-order,Rosvall-2014-memory,yin2017higher,Xu-2016-higher-order,Benson-2016-motif-spectral,Benson-2017-srw}. At a high-level, this constitutes an analysis of network of data in terms of multi-node patterns such as motifs and also in terms of stochastic processes that depend on more history. One of the origins of these studies is Milo's celebrated paper on the presence of higher-order interactions~\cite{milo_network_motifs}, which showed that some subgraphs appear more frequently than others. While there are plenty of network models (such as those mentioned above), the efficient ones often cannot model arbitrary higher-order interactions such as motifs and their interactions.

The primary contribution of this manuscript is a simple and flexible network model that has the ability to capture a single type of higher-order interaction. We call it the  HyperKron model. (This is introduced formally in \S\ref{sec:model}.) For arbitrary higher-order interactions, a more appropriate primitive is hypergraph modeling~\cite{Bollobas-2011-hypergraphs} to directly model the higher-order interactions---which is where our inspiration came from.  In comparison with many of the network models above, the key difference is that the probability model underlying it specifies a distribution on \emph{hyperedges} rather than edges. To generate a network, we then associate each hyperedge with a specific network motif (such as a triangle or feed-forward motif).  As we will show, this model exhibits clustering coefficients which can be closely aligned with real-world data. As might be guessed from the name, the model is a generalization of the extremely parsimonious Kronecker graph model~\cite{Leskovec-2005-Kronecker,leskovec2010_Kronecker,Seshadhri-2013-kronecker}.

One of the challenges with this model is that an exact and efficient sampling procedure for the desired hyperedge probability distribution is non-trivial to create. The Kronecker model, for instance, has historically been only approximately generated~\cite{moreno}. The situation is even more complex for the HyperKron model and we need to employ techniques including multidimensional Morton codes in order to generate these graphs in time proportional to the number of edges. (Our procedure is explained in \S\ref{sec:eff}). 

An advantage to working with the HyperKron model is it admits an analytical characterization of simple properties. We show, for instance, the number of hyperedges that share an edge (\S\ref{sec:dups}). This enables us to get accurate estimates of the number of edges the resulting model has for sparse graphs (\S\ref{sec:approxedges}). %We also empirically study emergent properties of the networks including XXX,XXX (Sections~\ref{sec:XX},~\ref{sec:XXX}).

We conclude the technical portion of our paper with case-studies on the HyperKron model. We show that the HyperKron model, when using a 3 node motif, generates substantial triadic clustering in fitting real-world network data, far beyond what is possible with Kronecker models (\S\ref{sec:clust}). We also illustrate the same generated graphs \emph{lack} clustering structure in four cliques that is present in real-world networks. We finally show that the model is flexible enough to model other types of interactions including directed and signed interactions when the motif associated with the hyperedge is one of the coherent-feed forward motifs (\S\ref{sec:flex}).

%\begin{enumerate}
%	\item Hypergraph modeling
%	\item Motifs
%	\item Graph modeling
%	\item Kronecker Graphs: They aren't good models, but non are
%	\item Propose HyperKron model: It has triangles, high degree skew, non-trivial clustering
%\end{enumerate}
%Graph models have long been used in lieu of real data which can be expensive and hard to come by. One such class of models constructs a matrix of probabilities, and samples an adjacency matrix by flipping a weighted coin for each entry. Examples include the Erdos-Reynii model~\cite{Erdos-1959-random}, Chung-Lu model~\cite{Chung-2002-random}, Stochastic Block Model~\cite{Holland-1983-sbm}, and the Kronecker Model~\cite{leskovec2010_Kronecker,leskovec2005,Chakrabarti-2004-rmat}. The efficiency of generating a graph in this way is not the topic of discussion here, but can be read about elsewhere \cite{Ramani-preprint-fast-graph-sampling}. 

%Here we present a model based closely off of the the Kronecker Model, but instead of placing edges we place triangles. 
%
%We also calculate expected feature counts for a number of useful features of the graph.

%Milo's study on network motifs~\cite{milo_network_motifs} \\
%Mixed Kronecker Graph Model~\cite{Moreno-2010-mKPMG}
\section{Preliminaries}
Our goal is to present the background, terminology, and notation to understand the HyperKron model presented in the next section. 

\textbf{Graphs and matrices.}
Let $G = (E,V)$ be an unweighted, undirected graph, where $V$ is the set of vertices and $E$ is the set of edges. Graphs can be represented by adjacency matrices $\mA$, where $\mA_{ij}$ is equal to 1 if $i$ and $j$ are connected by an edge, and 0 otherwise. $\mA$ is symmetric in an undirected graph, $\mA_{ij} = \mA_{ji}$. Extensions to directed and signed graphs are discussed in \S\ref{sec:flex}. The degree of a vertex $i$, denoted $d_i$ is the number of vertices, $j$, where $\mA_{ij} = \mA_{ji} = 1$.

\textbf{Sampling Graphs from Probability Matrices.}
There is a large body of work on modeling graphs, many of which were mentioned in the introduction and further detailed in \S\ref{sec:related}. What is relevant here is the class of graph generators which involves sampling edges from a probability matrix. Examples include the \ERfull model~\cite{Erdos-1959-random}, Chung-Lu model~\cite{Chung-2002-random}, Stochastic Block Model~\cite{Holland-1983-sbm}, and the Kronecker Model~\cite{leskovec2010_Kronecker,Leskovec-2005-Kronecker,Chakrabarti-2004-rmat}. This type of generator starts with a matrix of probabilities, $\mP$, with the number of rows and columns equal to the number of nodes desired in the graph. For each entry $i,j$ of $\mP$, set $\mA_{ij}$ equal to $1$ with probability $\mP_{ij}$, and set $\mA_{ij}$ equal to $0$ otherwise. This type of model allows for generating many instances of a graph from a single generator matrix $\mP$. 

\textbf{Kronecker Graph.} The HyperKron model which will be presented in \S\ref{sec:model} is built on many of the same motivations of the Kronecker Graph Model~\cite{leskovec2010_Kronecker,Leskovec-2005-Kronecker,Chakrabarti-2004-rmat}, so we briefly cover that model. Let $\mP$ be an $n \times n$ matrix of probabilities called an \emph{initiator} matrix, with $n$ small ($n$ between 2-5 is typical). The Kronecker Product of $\mP$ with itself is the $n^2 \times n^2$ matrix constructed by multiplying every entry of $\mP$ with itself. For example, if $\mP$ is the $2 \times 2$ matrix $\sbmat{ a & b \\ c & d }$, then the Kronecker product $\mP \otimes \mP$ is
 \[\mP^2 = \mP \otimes \mP = \begin{bmatrix} a\cdot \mP & b \cdot \mP\\ c \cdot \mP & d \cdot \mP \end{bmatrix} = \sbmat{ aa & ab & ba & bb \\ ac & ad & bc & bd \\ ca & cb & da & db \\ cc & cd & dc & dd  }. \]
Define the $r$th Kronecker product $\mP^{r}: n^r \times n^r$ to be $r$ Kronecker products of $\mP$ with itself:\footnote{It is worth being explicit here that we are abusing notation and use $\mP^r$ to indicate ``powering-up'' Kronecker products rather than the standard notation of ``powering-up'' by repeated matrix multiplication. We never multiply matrices in this paper and only multiply by Kronecker products.}
\[ \mP^r = \underbrace{\mP \otimes \mP \otimes \cdots \otimes \mP}_{r \text{ times}}.\]  Then  $\mP^r$ is the matrix of probabilities used to sample a graph. 

\textbf{Kronecker Graph Properties.} The Kronecker graph model became popular because of a number of desirable properties such as skewed degrees~\cite{Seshadhri-2013-kronecker} and similar properties to real-world networks~\cite{leskovec2010_Kronecker}. Additionally, storage of the initiator matrix $\mP$ is very cheap, at just $n^2$ entries (where $n$ is often taken to be 2). It has been used as a synthetic generator for parallel graph benchmarks~\cite{Murphy-2010-graph500} (the Graph500 benchmark). Choosing parameters in the Kronecker model to \emph{fit} a given graph has been studied using maximum-likelihood methods~\cite{leskovec2010_Kronecker} and method-of-moments estimators~\cite{Gleich-2012-kronecker}.

\textbf{Hyperedges.}
The HyperKron model will rely on the notion of \textit{hyperedges}. A hyperedge is just a set of vertices. It generalizes an edge which is just a set of two vertices. In this case all hyperedges in our model will have the same cardinality, which is often called a regular hypergraph. (For simplicity, we describe our model where each hyperedge has three vertices.) When we create a graph from a hypergraph in the HyperKron model, we associate each hyperedge with a motif. For most of our paper, this motif is simply a triangle (\S\ref{sec:model}, \ref{sec:eff}, \ref{sec:analytic}), which enables us to analyze some of the model properties. The HyperKron model is flexible enough to handle other hyperedge structures, as discussed in \S\ref{sec:flex}.  Though the HyperKron model uses the concept of hypergraphs, it is important to note that we are not concerned with the generation of hypergraphs. We simply use mechanisms to generate hyperedges to impose higher-order structure on a traditional graph.

\section{The HyperKron Model}
\label{sec:model}
The HyperKron Model mimics the ideas of the original Kronecker Model. The difference is that instead of starting with a matrix and Kronecker-powering it to get a large matrix of probabilities corresponding to edges, we start with a tensor and Kronecker-power it to get a massive tensor of probabilities corresponding to \emph{hyperedges}. For the sake of simplicity in our discussion and analysis, we consider hyperedges with up to three nodes (3d tensors) although the ideas extend beyond this setting. To generate a graph, we then associate the hyperedge with a triangle. The set-up extends to other motifs on three nodes as discussed in \S\ref{sec:flex}.

In more detail, we start with a 3 dimensional initiator tensor, $\cmP$, with dimensions $n \times n \times n$.  Just like in the Kronecker model, the value of $n$ should be small, between 2 to 5. For example take a $2 \times 2 \times 2$ symmetric initiator tensor:
\begin{equation}% 
\label{eq:sym-tensor-P}
	\includegraphics[width =0.2\linewidth]{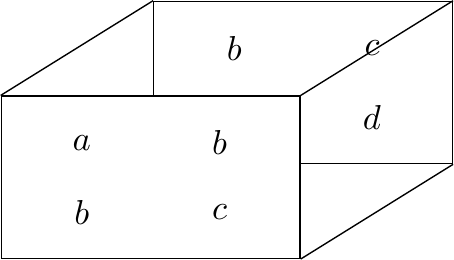}
\end{equation}%
Note that symmetric in the case of a tensor means any permutation of indices has the same value. $\cmP_{112} = \cmP_{121} = \cmP_{211}$. The entries of $\cmP$ should be probability values. (Again, our model is not restricted to symmetric tensors, it merely simplifies the exposition.)

\begin{marginfigure}
 \includegraphics[width=\linewidth]{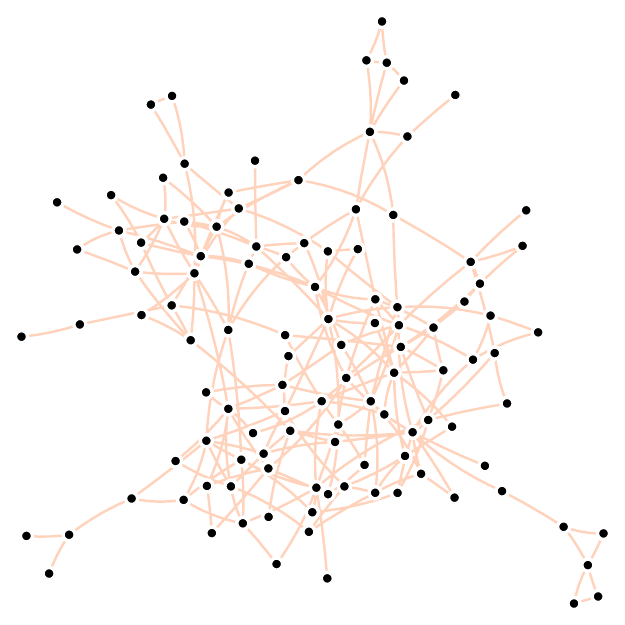}
  \caption{An example graph generated with the HyperKron model with symmetric initiator ($a \!=\! 0.8, b \!=\!0.115, c\!=\!0.215, d\!=\!0.61$), $r=7$ (only the large component is shown). Note the presence of individual edges not involved in triangles.}
  \label{fig:example}
\end{marginfigure}

The Kronecker product of tensors, $\cmP\otimes \cmP$, works just like the Kronecker product of matrices: every element gets multiplied by every other element giving way to a $n^2 \times n^2 \times n^2$ tensor~\cite{Phan-2012-kronecker,akoglu2008rtm}. In the example above, $\cmP^2 = \cmP\otimes \cmP$ has dimension $4 \times 4 \times 4$
\begin{equation}%
	\label{eq:sym-tensor-P2}
	\includegraphics[width=0.55\linewidth]{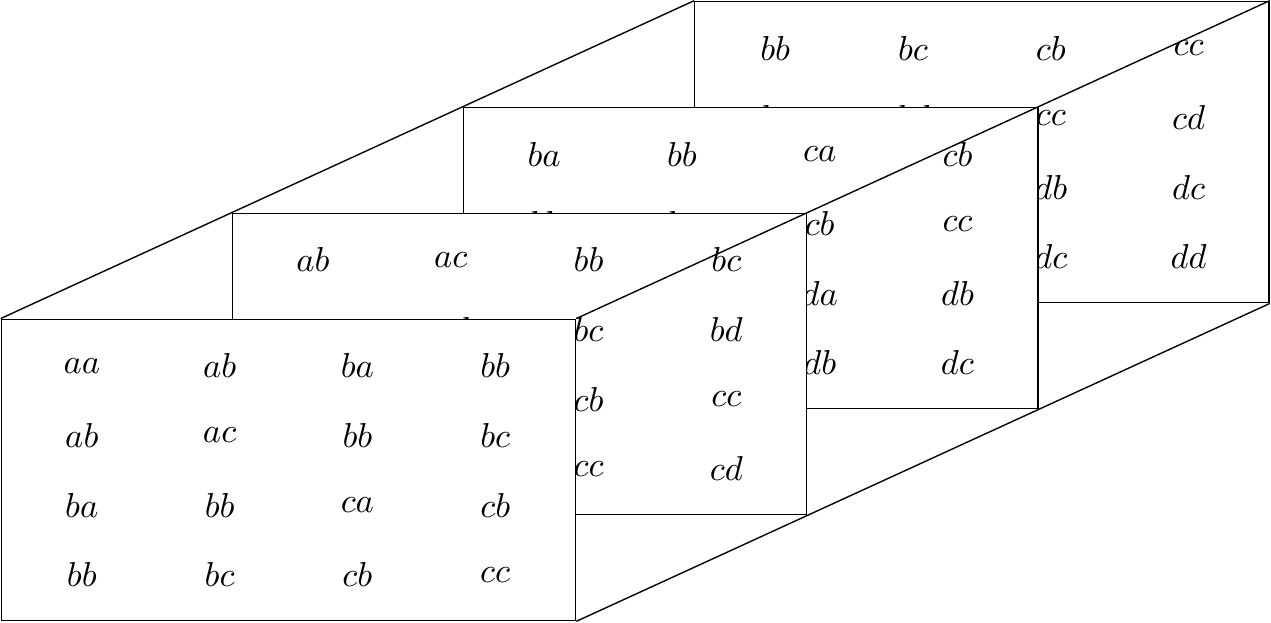}
\end{equation}%
% And the third Kronecker product, $\cmP^3$ has dimension $8 \times 8 \times 8$.
% \begin{equation}
% 	\includegraphics[width=0.8\linewidth]{3DMatrix_Kronned_2.pdf}
% \end{equation}
 \textbf{The Symmetric HyperKron Model with Triangles.} Given an $n \times n \times n$ initiator tensor $\cmP$ of probabilities, 
construct the $r$th Kronecker Product of $\cmP$, \[\cmP^r = \underbrace{\cmP \otimes \cmP \otimes \cdots \otimes \cmP}_{r \text{ times}}. \]
Then $\cmP^r$ is of dimension $n^r \times n^r \times n^r.$ Note that if $\cmP$ is symmetric then so is $\cmP^r$. Generate a set of hyperedges where we include hyperedge $(i,j,k)$ with probability $\cmP^r_{ijk}$. For each generated hyperedge, insert three undirected edges $(i,j)$, $(j,k)$, and $(i,k)$. Duplicate edges are coalesced into a single edge. This results in an undirected graph on $n^r$ vertices.  The values of $i,j,k$ need not be unique, so that we may just place an edge (or a loop). An example of the result is show in Figure~\ref{fig:example}. Because we insert undirected edges, it makes the most sense to consider this model with \emph{symmetric} tensors, in this case, we can restrict our generation to cases where $i \le j \le k$ (for instance) to minimize the number of duplicates. 
%We wish to sample a graph (adjacency matrix) using the matrix of probabilities $\cmP^r$. Note that $\cmP^r$ is 3-dimensional, but the graph will be 2-dimensional, $\mA: n^r \times n^r$. For each entry $(i,j,k)$ of $\cmP^r$, place triangle $(i,j,k)$ in $\mA$ with probability $\cmP_{ijk}$. That is, place three edges:  $(i,j), (j,k)$, and $(k,i)$ in the graph $\mA$. After placing all edges, remove loops and symmetrize to get the final graph. These details and others on efficiently generating $\mA$ are discussed in section \ref{sec:eff}.
	
%The model allows placing triangles with repeated indices, i.e. edges. This is realistic for a real world network, because we would expect some nodes to join the network with only 1 neighbor. That being said, in most cases, we expect that all nodes in the final graph are members of triangles. This is due to the fact that every edge $(i,j)$ has $n^k$ opportunities to be placed.

\section{Efficient Generation}
\label{sec:eff}
A simple algorithm to generate a HyperKron model is to explicitly generate the tensor $\cmP^r$ and then to explicitly sample Bernoulli random variables (coin-flips) for each entry in the tensor. If $N=n^r$ is the dimension of the tensor, this is an $O(N^3)$ algorithm, and will not enable us to efficiently generate realistically large networks. Consequently, we seek a more efficient algorithm. 

%This paragraph identifies us as the authors, so I've edited it below
%The ideal case for a generation algorithm is that we should do $O(m)$ or $O(m \log N)$ work where $m$ is the number of edges in the output. Note that $r = \log N$.  We will show how to get an $O(m r^2)$ method, which can be achieved by adapting the idea of \emph{grass hopping} from our recent paper on generating graphs from matrices of probabilities~\cite{Ramani-preprint-fast-graph-sampling}.  (This paper is written with a tutorial style and provides a gentle introduction to many of the topics we discuss in this section for those unfamiliar.) 

The ideal case for a generation algorithm is that we should do $O(m)$ or $O(m \log N)$ work where $m$ is the number of edges in the output. Note that $r = \log N$.  We will show how to get an $O(m r^2)$ method, which can be achieved by adapting the idea of \emph{grass hopping} from our recent paper on generating graphs from matrices of probabilities~\cite{Ramani-preprint-fast-graph-sampling}.  (This paper is written with a tutorial style and provides a gentle introduction to many of the topics we discuss in this section for those unfamiliar.) 

Historically, sampling Kronecker graphs was done using a ball-dropping approach~\cite{Leskovec-2005-Kronecker,Seshadhri-2013-kronecker} that inserted edges one at a time by simulating where a ball-dropped through successive initiator matrices would land. This gives an $O(mr)$ algorithm, at the cost of an approximate distribution. A faster method to generating an exact Kronecker graph was finally achieved when Moreno et al. showed that a Kronecker matrix is comprised of a small number of \ERfull regions~\cite{moreno} and illustrated a way to efficiently sample edges within these regions. 

Our approach is also based on the idea of looking at the \ERfull regions within the Kronecker-powered tensor $\cmP^r$. Recall that an \ERfull graph is sampled from a matrix where every edge has the same probability of occurring. An \ERfull region is a set of entries in $\cmP^r$ where all the probability values are the same. For instance, note that the probability $ab$ occurs multiple times in~\eqref{eq:sym-tensor-P2}. Edges in these regions can be generated by a waiting time, geometric variable, or grass-hopping method~\cite{FanMullerRezucha1962,BatageljBrandes2005,Hagberg2015, Ramani-preprint-fast-graph-sampling}. That is, we sample a geometric random variable to find the gap between successive edges. Thus, the method only does work proportional to the number of edges within the region.  What is difficult is to identify \emph{where} these \ERfull regions occur and then how to map from these regions back to entries in $\cmP^r$. 

In the remainder of this section, we show (i) that the number of \ERfull blocks is sufficiently small that this approach will work given that we have to at least \emph{look} at each block -- this analysis will show that the number of such blocks is the number of length $r$ multisets of integers $\{ 0, \ldots, n^3 -1 \}$; (ii) how to sample edges in a multiplication table view of the repeated Kronecker product by grass-hopping (that is, sampling geometric random variables) and unranking multiset permutations; and (iii) how to identify entries in $\cmP^r$ by translating the multiplication table indices through a Morton code procedure. The final algorithm is given in Figure~\ref{fig:algorithm} for reference. Note that our procedure discussed in this section assumes a general initiator tensor $\cmP$ that need not be symmetric. 

\begin{marginfigure}
	\centering
 \begin{minipage}{\linewidth}\scriptsize
 \noindent \textbf{HyperKron sampling algorithm} 
\begin{compactitem}[$\cdot$]
 \item For each length-$r$ multiset of $\{ 0, 1, \ldots, n^3-1 \}$ ({\footnotesize call it $s$}) 
 \item Compute the probability $p$ for region $s$
 \item Let $t$ be the total length of the region $s$
\item Set the index $I$ to $-1$ 
\item While the index $I$ is less than $t$ 
\begin{compactitem}[$\cdot$]
\item Sample a geometric random variable with prob $p$
\item  Increment the index $I$ by the sample
\item  If the index $I$ is still less than $t$
\begin{compactitem}[$\cdot$]
\item  Identify the multiset perm $p$ for $I$ 
\item  Compute the multiplication table index $J$ for $p$ 
%\hspace*{2em} $\cdot$ MortonDecode$(J,n,n,n)$ to get a hyperedge in $\cmP^r$
\item  MortonDecode$(J,n,n,n)$ gives a hyperedge in $\cmP^r$
\end{compactitem}
\end{compactitem}
\end{compactitem}
% \noindent \textbf{The HyperKron sampling algorithm.} \\
% $\cdot$ For each length-$r$ multiset of $\{ 0, 1, \ldots, n^3-1 \}$ ({\footnotesize call it $s$}) \\
% $\cdot$ Compute the probability $p$ for region $s$ \\
% $\cdot$ Let $t$ be the total length of the region $s$  \\
% $\cdot$ Set the index $I$ to $-1$ \\
% $\cdot$ While the index $I$ is less than $t$ \\
% \hspace*{1em} $\cdot$ Sample a geometric rand.~var.~with prob $p$ \\
% \hspace*{1em} $\cdot$ Increment the index $I$ by the sample \\
% \hspace*{1em} $\cdot$ If the index $I$ is still less than $t$\\
% \hspace*{2em} $\cdot$ Identify the multiset perm.~$p$ for $I$ \\
% \hspace*{2em} $\cdot$ Compute the multiplication table index $J$ for $p$ \\
% %\hspace*{2em} $\cdot$ MortonDecode$(J,n,n,n)$ to get a hyperedge in $\cmP^r$
% \hspace*{2em} $\cdot$ MortonDecode$(J,n,n,n)$ gives a hyperedge in $\cmP^r$
\end{minipage}
%\vspace*{-\baselineskip}
\caption{The pseudocode for our fast hyperedge sampling algorithm on a HyperKron model. An implementation is at \url{www.cs.purdue.edu/homes/dgleich/codes/hyperkron}
}
\label{fig:algorithm}
\end{marginfigure}

\subsection{A small number of \ERfull blocks}
\label{sec:multiset}
First, we show that there are fewer than $O(N)$ \ERfull regions in the graph. (If not, then this procedure would have trouble efficiently generating sparse graphs where $m = O(N)$.)
Let the initiator tensor $\cmP$ be $n\times n \times n$, and let $r$ be the number of HyperKron products, so that there are $N = n^r$ nodes in the graph.
Notice that each probability value in $\cmP^r$ is the product of $r$ values from $\cmP$. Entries of $\cmP$ can appear more than once in the product, and the entries of $\cmP^r$ are only unique up to permutation (that is, the multiplication of these probabilities from $\cmP$ is commutative). Thus, the total number of unique probability values in $\cmP^r$ is $n^3$ \textit{multi-choose} r:  
%\[  \multiset{n^3}{r} = \binom{n^3+r-1}{r} \]
%We need to be assured that the number of \ERfull regions of $\cmP^r$ is less than the number of nodes. If this weren't true, then visiting each \ERfull region would be more work than coin-flipping (visiting each node). 
%As $r$ becomes large, $\multiset{n^3}{r} = 
$\binom{n^3+r-1}{r}$. This goes to $O(r^{n^3-1})$, which is less than the $O(n^r)$ nodes of the graph. (This can be verified by a simple argument akin to one in~\citet{Ramani-preprint-fast-graph-sampling}).

\subsection{Multiplication tables \& HyperKron tensors}
\label{sec:multtables}
How can we distinguish each of the \ERfull regions? Each region is identifiable by its unique product of elements from $\cmP^r$, a probability value. Let us order these probability values. To that end, first number the entries of $\cmP$ 0 through $n^3-1$. Then when writing an entry of $\cmP^r$, associate each element in the product with the sequence of $r$ integers. For example, if $\cmP$ is $2 \times 2 \times 2$ as in~\eqref{eq:sym-tensor-P}, map each of the 8 entries to an index between 0 and 7. %(More generally, this is a map from $0$ to $n^3-1$.) 
Probability $ada$ in $\cmP^3$ would be mapped to $070$, where $a = 0$ and $d = 7$. 

It isn't obvious how to easily identify each of the locations of $ada$ in $\cmP^3$. (Or more generally, elements in $\cmP^r$.) We first solve an easier problem and then later determine how to translate back to $\cmP^r$. If we re-order the entries of $\cmP^3$ so that $ada = 070$ occurred exactly in locations $[0,0,7], [0,7,0], [7,0,0]$, then the locations would be very easy to find---they are just the permutations of $[070]$. 

We will call this re-ordering a \textit{multiplication table}. Define $v = \text{vec}(\cmP)$ to be the initiator tensor as a length-$n^3$ vector proceeding in a column-major fashion, e.g. the vectorized version of~\eqref{eq:sym-tensor-P} is $[a,b,b,c,b,c,c,d]$. Define a $r$-dimensional multiplication table: 
\begin{equation}
 M(\underbrace{i,j,\ldots,k}_\text{r \text{ indices}}) = \underbrace{v_i v_j \cdots v_k}_{r \text{ terms}}. 
\end{equation}
For instance, $M(0,0,7) = M(0,7,0) = M(7,0,0) = aad$. 
%\[ reshape(\underbrace{v \otimes v \otimes \cdots \otimes v}_{r \text{ times}} )\]
%Note that a HyperKron tensor $\cmP^r$ is always 3-dimensional, while a multiplication table is $r$-dimensional.

The start of our strategy is: for each unique probability in $\cmP^r$, given by a multiset of indices as in \S\ref{sec:multiset},  ``grass-hop'' sample through the locations in the multiplication table where the probability is all the same.  We will see how to do this efficiently next in \S\ref{sec:grass-hopping}, and finally see how to convert between entries of the multiplication table $M$ and $\cmP^r$ in \S\ref{sec:morton}.
%grass-hop through the permutations to find the locations for hyperedges in the multiplication table. Then find the corresponding locations in $\cmP^r$. At this point we need to establish $(1)$: how to grass-hop within an \ERfull region, and $(2)$ how map between between the Multiplication table and the Kronecker tensor.

\subsection{Grass-Hopping Kronecker Tensors}
\label{sec:grass-hopping}
Given an \ERfull region in $\cmP^r$ or $M$, we now discuss how to ``grass-hop'' within that region to find successive hyperedges. (For Kronecker graphs, see the treatment in~\citet{Ramani-preprint-fast-graph-sampling}).  Let us say that our \ERfull region corresponds with a probability which is mapped to indices $p = v(i_1) v(i_2) \cdots v(i_r)$ where $v = \text{vec}(\cmP)$ as described in \S\ref{sec:multtables}. Recall that each of the elements of $v$ are mapped to a numerical index between 0 and $n^3-1$. As established in \S\ref{sec:multtables}, the locations of $p$ in the $r$-dimensional multiplication table correspond exactly with permutations of $i_1, i_2, \ldots, i_r$. Note that these are permutations of \textit{multisets}, or sets in which elements can occur more than once. We label each permutation lexicographically from $0$ to $t-1$ where \[ t = \frac{m!}{a_1!a_2! \ldots a_r!} \]
is the number of permutations of the multiset $i_1, \ldots, i_r$, $m$ is the cardinality of the multiset, and $a_i$ is the number of times that the $i$th element appears.

The idea for generation then, is that we can easily identify indices between $0$ and $t-1$ where edges occur because each hyperedge occurs with the same probability $p$. As previously hinted, this is done by sampling a geometric random variable to compute the gap until the next edge (hence we ``grass-hop'' from edge-to-edge). See the discussions in~\citet{Ramani-preprint-fast-graph-sampling},~\citet{FanMullerRezucha1962},~\citet{BatageljBrandes2005}, and~\citet{Hagberg2015} for more about this technique.

Given the indices where hyperedges occurred, we then need to map them to entries of the multiplication table. This can be done by \emph{unranking} multiset permutations~\cite{knuth1997art,bonet}.
%We wish to order the multi-set permutations, in order to grass-hop among them. To determine this ordering on multiset permutations, we can use the ranking and unranking on combinatorial enumeration as described in~\cite{bonet}.

For example, suppose that the \ERfull region corresponds to a probability with indices $[0,1,1,2]$. The permutations of this multiset are (lexicographically):
\begin{smalleq}%
\begin{equation}
\begin{array}{llll}
\left[ 0,1,1,2 \right] \rightarrow 0 & \left[0,2,1,1 \right] \rightarrow 2 & \ldots & \left[2,2,0,1 \right] \rightarrow 10 \\
 \left[0,1,2,1 \right] \rightarrow 1 & \ldots & \ldots &  \left[2,2,1,0 \right] \rightarrow 11 
\end{array} 
 \end{equation}
\end{smalleq}% 
The unranking of this multiset corresponds to taking one of the indices $0, \ldots, 11$ and generating the corresponding permutation. This step can be done in time $O(r^2)$ without any precomputation. 
The process of finding the index associated with a permutation of a multiset is called \textit{ranking}. The inverse process, finding the permutation given an index, is called \textit{unranking}. We omit the details of this process, it is not difficult to understand, and can be found in our online code and in~\citet{Ramani-preprint-fast-graph-sampling}.

%We now can use a geometric random variable to repeatedly "hop" to the next successful hyperedge, through the labels 0-(number of permutations -1), until the end is reached. This gives a list of entries in the Multiplication table.

\subsection{Morton Codes}
\label{sec:morton}
The last detail is how to map between the $r$-dimensional multiplication table entry and the 3-dimensional HyperKron tensor $\cmP^r$. The relationship between their locations depends on \textit{Morton codes} as was described for the case of matrices~\cite{Ramani-preprint-fast-graph-sampling}. We extend that analysis to the 3-dimensional HyperKron tensor. A Morton code reflects a particular way of ordering multidimensional data. Because of its shape for a matrix, it is sometimes called a ``Z-order.'' For a 3d space (such as a tensor), the order is given by a more complicated structure illustrated in Figure~\ref{fig:3DMorton}. The particular relationship we use is established by the following theorem.

\begin{theorem}
	Let $\cmP^r$ be an $n\times n \times n$ tensor, and $\vv$ be the column major representation of $\cmP$. Consider an element in the vectorized multiplication table $M$ 
	 \[\underbrace{\vv \otimes \vv \otimes \ldots \otimes \vv}_{r \text{ times}} \] with index $(p_1, p_2, \ldots, p_r)$. Let $I$ be the lexicographical index of the element $(p_1,p_2,\ldots,p_r)$. Then the base 3-dimensional Morton Decoding of $I$ in base $n$ provides the row, column, and slice indices of an element in \[ \underbrace{\cmP \otimes \cmP \otimes \ldots \otimes \cmP}_{r \text{ times}} \] with the same value. 
\end{theorem}
\emph{Proof idea.} The full proof of this result is tedious and uninsightful, so we omit it in the interest of space. See~\citet{Ramani-preprint-fast-graph-sampling} for the proof for the 2-dimensional Morton decoding. The key idea is we can induct over $r$ and show that given that it works for $r=1$ (trivially) and given that it works for $(p_1, \ldots, p_{r-1})$, then the Morton code including $p_r$ will ``zoom in'' on the right region of the full tensor $\cmP^r$. 

Given an index $I$ from the $(n^3)^r$ indices of the multiplication table, the 3d Morton decoding procedure works by first expressing $I$ in base $n$. Note, $I$ will have at most $3r$ digits in base $n$. Then given the digit expansion of $I$ in base $n$: $I = (d_1, d_2, d_3, d_4, \ldots, d_{3r})_n$, the row in the tensor $\cmP^r$ is given by the number $R = (d_1, d_4, d_7, \ldots, d_{3r-2})_n$, the column in the tensor is given by $C = (d_2, d_5, \ldots, d_{3r-1})_n$, and the slice is given by $S = (d_3, d_6, \ldots, d_{3r})_n$. That is, we extract alternating sets of digits from $I$ to find the row, column, and slice. 
%Suppose $\cmP$ is a $3$-by-$3$-by-$3$ initiator tensor, and $r = 4$. Suppose that in the $4$-dimensional multiplication table we generate a hyperedge with linear index 329. The value of 329 in base 4 is $12011_4$. The resulting Morton Decode is

\begin{tuftefigure}
\centering
		\includegraphics[width = 0.2\textwidth]{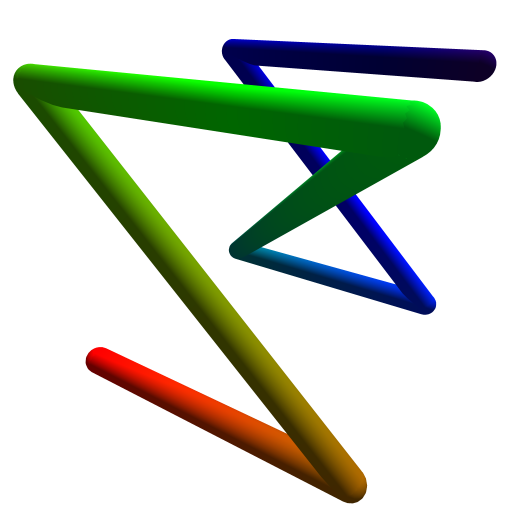}
		\includegraphics[width = 0.2\textwidth]{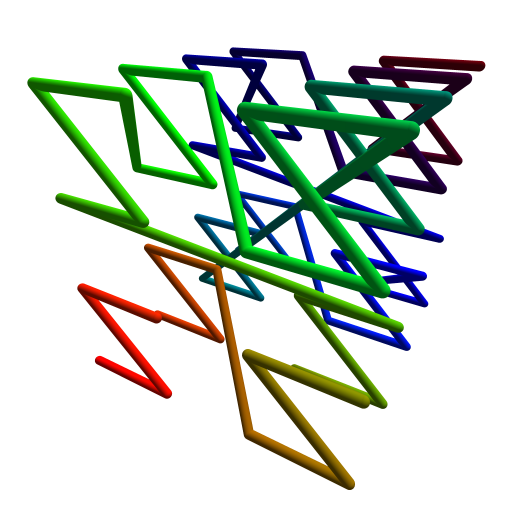}
		\includegraphics[width = 0.2\textwidth]{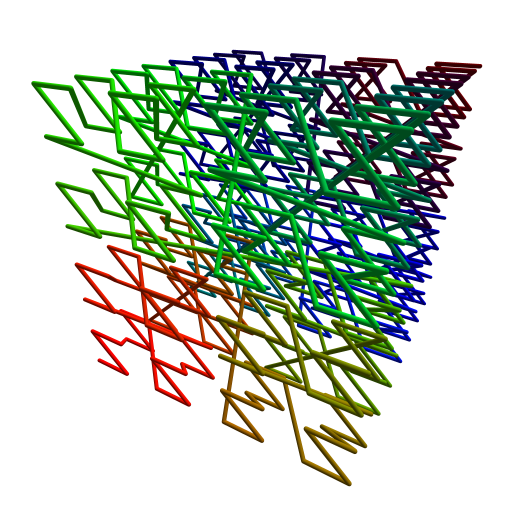}
\caption{3-dimensional Morton Codes. Starting from the left, a $2 \times 2 \times 2$ tensor, a $4 \times 4 \times 4$ tensor, and an $8 
	\times 8 \times 8$ tensor. Image obtained from~\citet{HoedtMorton} and used with permission.}
\label{fig:3DMorton}
%\vspace*{-\baselineskip}
\end{tuftefigure}

\subsection{Runtime performance}
\label{sec:runtime}

\begin{marginfigure}[-18\baselineskip]
 \includegraphics[width=\linewidth]{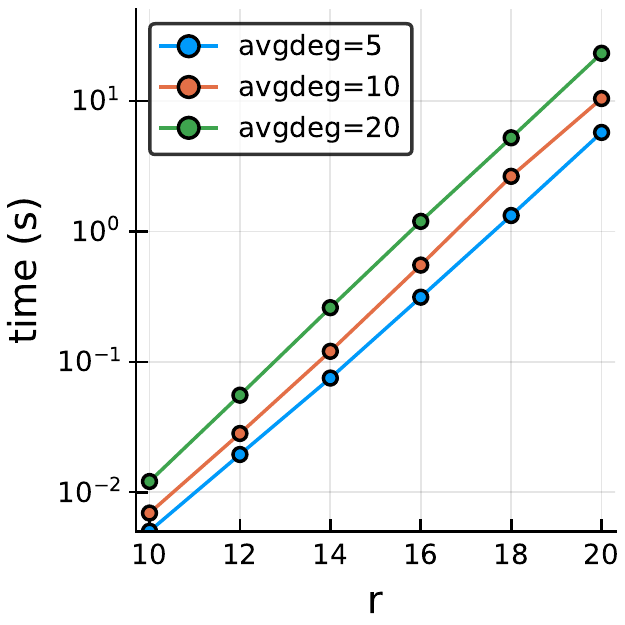}
  \caption{Time taken to generate hyperedges for a HyperKron model as $r$ varies shows linear scaling in $2^r$. The \emph{average degree} is the expected number of hyperedges per node. See \S\ref{sec:runtime} for more about the choice of parameters.}
  \label{fig:runtime}
\end{marginfigure}

We implemented the generation procedure (Figure~\ref{fig:algorithm}) in the Julia language with a goal towards optimizing easy-to-avoid computational overhead. The resulting program, 
%which is available from \url{www.cs.purdue.edu/homes/dgleich/codes/hyperkron}, 
which will be provided online upon acceptance,
generates hyperedges at about 1,000,000 per second on a single-thread of modern desktop computer. We evaluated the scalability of the code up to 20 million edges in three scenarios. All three scenarios use a $2 \times 2 \times 2$ symmetric initiator tensor $\cmP$. This gives four parameters $a,b,c,d$ as in equation~\eqref{eq:sym-tensor-P}. 
In the first case, we set $a = 0.05, b = 0.3, c = 0.4$, and choose $d$ such that the expected number of hyperedges ($(a + 3b + 3c + d)^r$) is $5$ times the number of nodes. (For reference, when $r=10$ then $d = 0.199$ and when $r=20$ then $d=0.018$.)  In the second case, we set $a = 0.9, b = 0.3, d = 0.0$ and choose $c$ such that the expected number of hyperedges is 10 times the number of nodes. In the third case, we set $a = 0.3, c = 0.3, d=0.1$ and choose $b$ so there are 20 times the number of hyperedges as nodes. The time it takes to generate graphs as $r$ varies from $10$ to $20$ is shown in Figure~\ref{fig:runtime}. Although the theoretical scaling of our procedure is $O(m r^2)$, we observe linear scaling in this regime because the $r^2$ work can be done efficiently within an array of $4r$ bytes that typically fits in $L1$ cache. Hence, it operates faster than many other steps of the algorithm.

\section{HyperKron Properties}
\label{sec:analytic}
Next we discuss analytic results and features of the HyperKron model. We compute various expected properties of the model such as the number of hyperedges that share an edge and practical estimates of the total number of generated edges. We also show that the model has non-trivial clustering for a wide set of parameters. %n section \ref{sec:fitting} clustering patterns which can be modeled in the HyperKron model .
%\begin{figure}
%\begin{center}
%\includegraphics[width=0.4\textwidth]{features.pdf}
%\end{center}
%\caption{Graph features that we can count in the HyperKron model.}
%\label{fig:features}
%\end{figure}
\subsection{Summation Formulas}
\label{sec:formulas}
Many of the quantities we analytically compute require the following summation formulas. Define ${\sum}^{\ast}$ to be a summation over all combinations of indicated indices except when two or more have the same value. While it will be easy to write down formulas relating to the model in terms of ${\sum}^{\ast}$, it is easier to sum over the original $\sum$. So here we present formulas that relate the two. Further explanation for these formulas can be found in \citet{Gleich-2012-kronecker}. 
%The last formula for the sum over 5 indices can be found in a similar description as described in \cite{Gleich-2012-kronecker} for the sum over 4 indices. Define $g_{ijkl} = \sum_m f_{ijklm} - f_{ijkli} - f_{ijklj} - f_{ijklk} - f_{ijkll}$. Then take ${\sum}^{\ast}_{ijklm} f_{ijklm} = {\sum}^{\ast}_{ijkl} g_{ijkl}$ and use the formula over 4-sums.
%\begin{smalleq}%
\[ {\sum_{ij} }^{\ast}f_{ij} = \sum_{ij} f_{ij} - \sum_i f_{ii}  \]
\[  {\sum_{ijk}}^{\ast} f_{ijk} = \sum_{ijk} f_{ijk} - \sum_{ij} (f_{ijj} + f_{iji} + f_{iij}) + 2 \sum_i f_{iii} \]
%\end{smalleq}%

%\[ {\sum_{ijk_1k_2}}^{\ast} f_{ijk_1k_2} = \sum_{ijk_1k_2} f_{ijk_1k_2} - \sum_{ijk}( f_{ijki} + f_{ijkj} + f_{ijkk} + f_{ijik} + f_{ijjk} + f_{iijk}) \]
%\vspace{-.05in}
%\[\hspace{.5in}+ \sum_{ij} [ 2(f_{ijjj} + f_{ijii} + f_{iiji} + f_{iiij}) + f_{ijij} + f_{ijji} + f_{iijj} ]- 6\sum_if_{iiii} \]

%
%\begin{eqnarray}
%{\sum}^{\ast}_{ijklm} f_{ijklm} &=& \sum_{ijklm} f_{ijklm} - \sum_{ijkl} (f_{ijkli} + f_{ijklj} + f_{ijkll} + f_{ijkil} + f_{ijkjl} + f_{ijkkl} + f_{ijjkl} + f_{iijkl})\nonumber \\
%&&+\,\, \sum_{ijk} [2(f_{ijkii} + f_{ijkjj} + f_{ijkkk} + f_{ijiki} + f_{ijjkj} + f_{iijki} + f_{ijjjk} + f_{ijiik} + f_{iijik} + f_{iiijk}) \nonumber \\
%&&+\,\, f_{ijkij} + f_{ijkik} + f_{ijkji} + f_{ijkjk} + f_{ijkki} + f_{ijkkj} + f_{ijikj} + f_{ijikk} + f_{ijjki} + f_{ijjkk} + f_{iijkj} \nonumber \\
%&&+\,\, f_{iijkk} + f_{ijijk} + f_{ijjik} + f_{iijjk}] - \sum_{ij}[ 3(f_{ijjjj} + f_{ijiii} + f_{iijii} + f_{iiiji} + 2f_{iiiij}) \nonumber \\
%&&+\,\, 2(f_{ijiji} + f_{ijijj} + f_{ijjii} + f_{ijjij} + f_{iijji} + f_{iijjj}) + f_{ijjji} + f_{ijiij} + f_{iijij} + f_{iiijj}   ] \nonumber \\
%&&+\,\, 24\sum_{i} f_{iiiii} \nonumber
%\end{eqnarray}

\subsection{Exact Expectation of Edges}
\label{sec:edges}
First note that the actual number of edges is
\begin{equation}
 E = \frac{1}{2} {\sum_{ij}}^{\ast} \mA_{ij}
\end{equation}
 where $\mA$ is the symmetric adjacency matrix and ${\sum}^{\ast}$ is defined in \S\ref{sec:formulas}. 
Let $F_{ijt} = \text{Bernoulli}(\cmP^r_{ijt})$, so $F_{ijt}$ is 1 with probability $\cmP^r_{ijt}$ and 0 otherwise. The probability that $\mA_{ij} = 1$ is equal to the probability that \emph{any} of the Bernoulli samples $F_{rst}$ contain the indices $i$ and $j$. 

Now define a random variable, $X_{ij}$, for the number of successful trials of $F_{ijt}$ for all $t$ and permutations. Note $X_{ij} \in \{0,1,2,\cdots \}$, and we can calculate the probability of edge $(i,j)$ as the probability that $X_{ij}$ is greater than or equal to 1.
%$X_{ij} = \sum_{t} (F_{ijt}  +F_{itj} + F_{jit} + F_{jti} + F_{tij} + F_{tji}  ) = 6 \sum_t F_{ijt}$. Then we can calculate the expected value for each entry of $\mA$.
%\[  \mathbb{E}[X_{ij}] = 6 \sum_t \mathbb{E}[F_{ijt}] = 6 \sum_{t = 1}^n P_{ijt}   \]
%\begin{smalleq}%
\[ 
%\begin{aligned}
\mathbb{E}(\mA_{ij}) = \mathbb{P}(X_{ij} \geq 1) = 1- \mathbb{P}(X_{ij} = 0) =  1 - \prod_{t = 1}^n \mathbb{P}(F_{ijt} = 0) 
 =  1 - \prod_{t =1}^n (1 - \cmP^r_{ijt}) 
%\end{aligned}
\]
%\end{smalleq}%
Then the expected number of edges in the graph is the sum over all entries in the adjacency matrix. We divide by two since the summation counts each edge twice.
%\begin{smalleq}%
\[ \begin{aligned}
2\mathbb{E}(E) & = {\sum_{ij}}^{\ast} \mathbb{E}(\mA_{ij}) = {\sum_{ij}}\bigl[  1 \!-\! \prod_{t =1}^n (1\!-\! \cmP^r_{ijt})\bigr]  - {\sum_{i}}\bigl[  1 \!-\! \prod_{t =1}^n (1\!-\! \cmP^r_{iit})\bigr]
\end{aligned}\] 
%\end{smalleq}
%\begin{smalleq}%
%\begin{eqnarray}
%%E &=& \frac{1}{2} {\sum_{ij}}^{\ast} \mA_{ij} \nonumber \\
%2\mathbb{E}(E) &=& {\sum_{ij}}^{\ast} \mathbb{E}(\mA_{ij}) \nonumber \\
% &=& {\sum_{ij}}^{\ast} \left[  1 - \prod_{t =1}^n (1- \cmP^r_{ijt})\right] \nonumber \\
% &=& {\sum_{ij}}\left[  1 - \prod_{t =1}^n (1- \cmP^r_{ijt})\right]  - {\sum_{i}}\left[  1 - \prod_{t =1}^n (1- \cmP^r_{iit})\right] \nonumber 
%\end{eqnarray}
%\end{smalleq}
The very last equality comes from  \S\ref{sec:formulas}.
This formula, unfortunately, is not computationally helpful.

\subsection{Duplicate Edges Motif}
\label{sec:dups}
Next we explicitly compute the expected number of duplicate edges placed in the HyperKron model, which we will put to use for a computationally efficient approximation to the number of edges in \S\ref{sec:approxedges}. We expect to see duplicates if two hyper-edges are dropped with a single repeated edge. The number of such features in the model is the sum over all hyper-edges which share an edge, ${\sum_{k_1k_2}}^{\ast}({\sum_{ij}}^{\ast}\cmP^r_{ijk_1} \cmP^r_{ijk_2})$. We split the sums in this way because $k_1 \neq k_2$ and $i \neq j$, but other equalities among indices can occur. Using the relationships in \S\ref{sec:formulas},
%\begin{smalleq}
	\[
	\begin{aligned}
	& 4 \cdot \text{duplicates} =  {\sum_{k_1k_2}}^{\ast}({\sum_{ij}}^{\ast}\cmP^r_{ijk_1} \cmP^r_{ijk_2}) 
	  = {\sum_{ijk_1k_2}}\cmP^r_{ijk_1} \cmP^r_{ijk_2}  
	   \quad - \sum_{ik_1k_2} \cmP^r_{iik_1} \cmP^r_{iik_2} - \sum_{ijk_1 } (\cmP^r_{ijk_1})^2 + \sum_{ik_1} (P_{iik_1})^2 
	\end{aligned}
	\]
%\end{smalleq}
The factor of 4 is due to counting $\{\{ijk_1\}, \{ijk_2\} \}$ 4 times.

We can derive formulas for each of these sums in terms of the values of $\cmP$. The formulas will be based off of a $2 \times 2 \times 2$ non-symmetric initiator matrix, hence is more general than the HyperKron paradigm we've presented. It is easily adjusted to the symmetric case.
%\begin{center}
%	\includegraphics[width = 0.2\textwidth]{3DMatrix.pdf}
%\end{center}
We will show how to derive the summation for arguably the most difficult of those presented here, due to the 4 summation indices. The others are similar and we give their values in Eq.~\eqref{eq:formulas}.
\begin{lemma}%
	Let the entries of the initiator tensor $\cmP$ be
	%\begin{smalleq}%
	\begin{equation}%
		\begin{array}{llll}%
		\cmP_{111} = a_1 &  \cmP_{121} = a_2& \cmP_{211} = a_3 & \cmP_{221} = a_4 \\
		\cmP_{112} = b_1 & \cmP_{122} = b_2 & \cmP_{212} = b_3 &  \cmP_{222} = b_4 \nonumber
		\end{array}, \qquad \text{then}%
		\end{equation}%
	%\end{smalleq}% 
	%\begin{smalleq}%
	\[ \sum_{ijk_1k_2} \cmP^r_{ijk_1} \cmP^r_{ijk_2} = [(a_1 + b_1)^2 + (a_2 + b_2)^2 + (a_3 + b_3)^2 + (a_4+ b_4)^2]^r \]
	%\end{smalleq}%
\label{lemma:dups}
\end{lemma}%
\textsc{Proof.} We proceed inductively. 
	If $r = 1$, then $\cmP^r = \cmP$ is the initiator tensor, then the indices $i,j,k_1,k_2$ all range between 1-2, and we can easily write out the sum:
	%\begin{smalleq}%
	\[ \sum_{ijk_1 k_2} \cmP_{ijk_1} \cmP_{ijk_1 k_2}
		= (a_1 + b_1)^2 + (a_2 + b_2)^2 + (a_3 + b_3)^2 + (a_4 + b_4)^2 \]
		%\end{smalleq}%
	Assume the formula for $r -1$, 
%	\begin{smalleq}%
%		\[\sum_{ijk_1 k_2} \cmP^{r-1}_{ijk_1} \cmP^{r-1}_{ijk_2}  = [(a_1 + b_1)^2 + (a_2 + b_2)^2 + (a_3 + b_3)^2 + (a_4 + b_4)^2]^{r-1},\]
%	\end{smalleq}%
    we use $\cmP^r = \cmP \otimes \cmP^{r-1}$ to get: 
	%\begin{smalleq}%
	\[ \begin{aligned}
		& \sum_{ijk_1 k_2} \cmP_{ijk_1}^r \cmP_{ijk_2}^r = \sum_{ijk_1 k_2} \left(\cmP_{ijk_1} \otimes \cmP_{ijk_1}^{r-1} \right) \left(\cmP_{ijk_2} \otimes \cmP_{ijk_2}^{r-1} \right) \nonumber \\
		& \quad \quad \quad  = \sum_{ijk_1k_2} \cmP_{ijk_1} \cmP_{ijk_2}  \left( \sum_{ijk_1k_2} \cmP_{ijk_1}^{r-1} \cmP_{ijk_2}^{r-1}  \right) \nonumber \\
		&\quad \quad \quad = [(a_1 + b_1)^2 + (a_2 + b_2)^2 + (a_3 + b_3)^2 + (a_4 + b_4)^2] \nonumber \\
		&\quad \quad \quad \quad  \cdot [(a_1 + b_1)^2 + (a_2 + b_2)^2 + (a_3 + b_3)^2 + (a_4 + b_4)^2]^{r-1} \nonumber \\
		& \quad \quad \quad = [(a_1 + b_1)^2 + (a_2 + b_2)^2 + (a_3 + b_3)^2 + (a_4 + b_4)^2]^{r} \nonumber \qed \\
		\end{aligned}%
		\]%
	%\end{smalleq}%
%

\subsection{Approximate Expectation of Edges}
\label{sec:approxedges}
The formula for edges presented in \S\ref{sec:edges} is exact in expectation, but it is computationally \emph{expensive}. We instead offer an approximation for the number of edges that is appropriate when there aren't too many hyperedges. Our estimate comes from the basic idea that 3 times the number of hyper-edges dropped in the model, with small adjustments, should be a good estimate on the total number of edges assuming a sparse set of hyperedges. So to be more precise, our estimate is 3 times the number of 3-edges dropped \emph{plus} 2 times the number of two-edges dropped \emph{minus} duplicates expected at random (\S\ref{sec:dups}).

The number of 3-edges dropped is number of hyper-edges dropped with unique indices. This quantity is just the sum over $\cmP^r$:
%\begin{smalleq}%
\[ 6\cdot \text{3-edges} = {\sum_{ijk}}^{\ast} \cmP^r_{ijk} 
= \sum_{ijk} \cmP^r_{ijk} - \sum_{ij}( \cmP^r_{ijj} + \cmP^r_{jij} + \cmP^r_{jji} ) + 2\sum_{i} \cmP^r_{iii} \]
%\end{smalleq}%
The factor of 6 is because there are six permutations of $(i,j,k)$. 
%The formula uses the relationship between ${\sum}^{\ast}$ and $\sum$ discussed in section \ref{sec:formulas}. 

The number of 2-edges dropped is the number of hyper-edges dropped with a repeated index, which is exactly the end pieces of the formula above: 
%\begin{smalleq}%
\[ 
2 \cdot \text{2-edges} =  {\sum}^{\ast}_{ij}(\cmP^r_{ijj} + \cmP^r_{jij}+ \cmP^r_{jji}) \nonumber=\sum_{ij}( \cmP^r_{ijj} + \cmP^r_{jij} + \cmP^r_{jji} ) - 3\sum_{i} \cmP^r_{iii} 
\] 
%\end{smalleq}%
The factor of 2 is for counting $(i,j)$ twice.

%\begin{figure}
%	\begin{minipage}{0.2\linewidth}
%		\includegraphics[width=\linewidth]{dup.pdf}
%	\end{minipage}~
%	\begin{minipage}{0.45\linewidth}
%		\caption{Two hyperedges which share an edge.}
%		\label{fig:dup}
%	\end{minipage}
%\end{figure}

One of the sums from the duplicates formula is explicitly calculated in \S\ref{sec:dups}, and given the same setup as in lemma \ref{lemma:dups} the remaining sums can be found in similar fashion:
%\begin{smalleq}%
	\begin{equation} \label{eq:formulas}
	\begin{array}{l}%
	\sum_{ijk} (\cmP^r_{ijk})^m = (a_1^m + a_2^m + a_3^m + a_4^m + b_1^m + b_2^m + b_3^m + b_4^m)^r  \\
	%\sum_{ijk} \cmP^r_{ijk} \cmP^r_{iji} =  (a_1^2 + a_1^{}b_1^{} + a_2^2 + a_2^{}b_2^{} + b_3^2 + a_3^{}b_3^{} + a_4^{}b_4^{} + b_4^2)^r  \\
	%\sum_{ijk} \cmP^r_{ijk} \cmP^r_{ijj} =  (a_1^2 + a_1^{}b_1^{} + b_2^2 + a_2^{}b_2^{}+ a_3^2 + a_3^{}b_3^{} + a_4^{}b_4^{} + b_4^2)^r  \\
	\sum_{ijk} \cmP^r_{iij} \cmP^r_{iik} = ((a_1 + b_1)^2 + (a_4 + b_4)^2)^r  \\
	% \sum_{ij} \cmP^r_{iij} \cmP^r_{iii} = (a_1^2 + a_1^{}b_1^{} + a_4^{}b_4^{} + b_4^2 )^r  \\
	%\sum_{ij} \cmP^r_{iji} \cmP^r_{ijj} = (a_1^2 + a_2^{}b_2^{} + a_3^{}b_3^{} + b_4^2 )^r \\
	 \sum_{ij} (\cmP^r_{ijj})^m = (a_1^m + b_2^m + a_3^m + b_4^m)^r \\
	\sum_{ij} (\cmP^r_{iji})^m = (a_1^m + a_2^m + b_3^m + b_4^m)^r  \\ \sum_{ij} (\cmP^r_{iij})^m = (a_1^m + b_1^m + a_4^m + b_4^m)^r \\
	\sum_{i} (\cmP^r_{iii})^m = (a_1^m + b_4^m)^r
	\end{array} 
	\end{equation}
%\end{smalleq}% 
So all together, the estimate for the number of edges is 
%\begin{smalleq}%
\begin{equation}
\begin{aligned}%
\label{eqn:expectededges}
\mathbb{E}(E)&= 3(\text{3-hyperedges}) + 2(\text{2-hyperedges}) - \text{duplicates}  \\
%&=& 3\frac{1}{6}\left(\sum_{ijk} \cmP^r_{ijk} - \sum_{ij}( \cmP^r_{ijj} + \cmP^r_{jij} + \cmP^r_{jji} ) + 2\sum_{i} \cmP^r_{iii} \right) \nonumber \\
%&&+ 2\frac{1}{2} \left(\sum_{ij}( \cmP^r_{ijj} + \cmP^r_{jij} + \cmP^r_{jji} ) - 3\sum_{i} \cmP^r_{iii} \right) \nonumber \\
%&&- \frac{1}{4} ( \sum_{ijk_1k_2} \cmP^r_{ijk_1} \cmP^r_{ijk_2} \nonumber  - \sum_{ijk} ( 2\cmP^r_{ijk} \cmP^r_{iji} + 2 \cmP^r_{ijk} \cmP^r_{ijj} + (\cmP^r_{ijk})^2  \nonumber \\
%&&+ \cmP_{iij} \cmP_{iik} ) + \sum_{ij} ( 2 (\cmP^r_{ijj})^2 + 2(\cmP^r_{iji})^2 + 4\cmP^r_{iij}\cmP^r_{iii} + 2\cmP^r_{iji} \cmP^r_{ijj} \nonumber\\
%&& + (\cmP^r_{iij})^2 ) - 6\sum_{i} (\cmP^r_{iii})^2 ) \nonumber \\
%&=& \frac{1}{2}\sum_{ijk} \cmP^r_{ijk} +\frac{1}{2} \sum_{ij}( \cmP^r_{ijj} + \cmP^r_{jij} + \cmP^r_{jji} ) - 2\sum_{i} \cmP^r_{iii} \nonumber \\
%&&- \frac{1}{4} ( \sum_{ijk_1k_2} \cmP^r_{ijk_1} \cmP^r_{ijk_2} \nonumber  - \sum_{ijk} ( 2\cmP^r_{ijk} \cmP^r_{iji} + 2 \cmP^r_{ijk} \cmP^r_{ijj} + (\cmP^r_{ijk})^2  \nonumber \\
%&& + \cmP_{iij} \cmP_{iik} ) + \sum_{ij} ( 2 (\cmP^r_{ijj})^2 + 2(\cmP^r_{iji})^2 + 4\cmP^r_{iij}\cmP^r_{iii} + 2\cmP^r_{iji} \cmP^r_{ijj}  \nonumber\\
%&& + (\cmP^r_{iij})^2 ) - 6\sum_{i} (\cmP^r_{iii})^2 ) \nonumber \\
&= 1/2(a_1 + a_2 + a_3 + a_4 + b_1 + b_2 + b_3 + b_4)^r  
 +1/2(a_1 + b_2 + a_3 + b_4)^r \\
 & + 1/2(a_1 + a_2 + b_3 + b_4)^r  
  +1/2 (a_1 + b_1 + a_4 + b_4)^r -2 (a_1 + b_4)^r  \\
 &- 1/4 \left( (a_1 + b_1)^2 + (a_2 + b_2)^2 + (a_3 + b_3)^2 + (a_4 + b_4)^2 \right)^{r}   \\
& + 1/4(a_1^2 + a_2^2 + a_3^2 + a_4^2 + b_1^2 + b_2^2 + b_3^2 + b_4^2)^r  \\
 & + 1/4 ((a_1 + b_1)^2 + (a_4 + b_4)^2)^r -1/4 (a_1^2 + b_1^2 + a_4^2 + b_4^2)^r  \\
\end{aligned}%
\end{equation}%
%\end{smalleq}%

To verify that the estimate \eqref{eqn:expectededges} is accurate, we test the expected edges count against the true number of edges generated in HyperKron models with the same parameters used as in the experiment in \S\ref{sec:runtime}. The true number of edges compared to the estimate is presented in Figure~\eqref{fig:edges}.  If you look closely, particularly for small $r$, \eqref{eqn:expectededges} is not exact. Nevertheless, it is quite accurate. It is important to note that \eqref{eqn:expectededges} will not work for all choices of parameters, particularly those that result in \textit{dense} graphs. For example, choosing parameters $a,b,c,d = (0.99,0.43,0.4,0.009)$ with $r = 13$ leads to an average of 4 million edges generated in a HyperKron graph, while the expected edge count from \eqref{eqn:expectededges} is only $1.98$ million. Nonetheless, it is still very useful when fitting sparse real-data to have a way to predict the number of edges which will appear in the  HyperKron model.

\subsection{Non-trivial Clustering}
\label{sec:non-trivial}
\begin{marginfigure}[-35\baselineskip]
	\includegraphics[width=\linewidth]{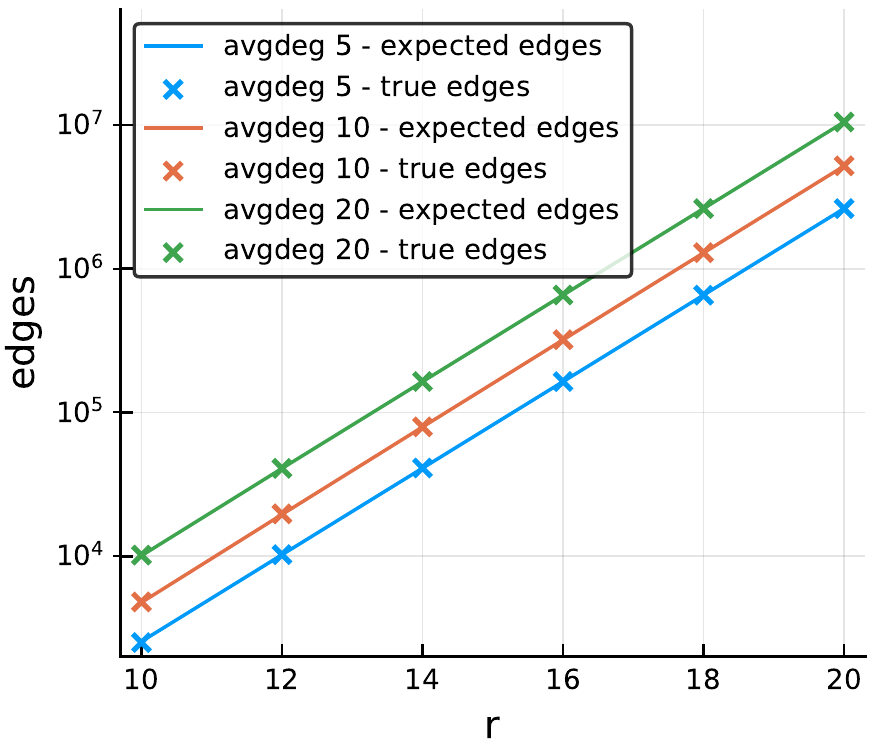}
	%\vspace*{-2.0\baselineskip}
	\caption{The number of edges generated in the HyperKron model, versus the count from \eqref{eqn:expectededges}. See \S~\ref{sec:runtime} for more about the choice of parameters and average degree.}
	\label{fig:edges}
	\includegraphics[width=\linewidth]{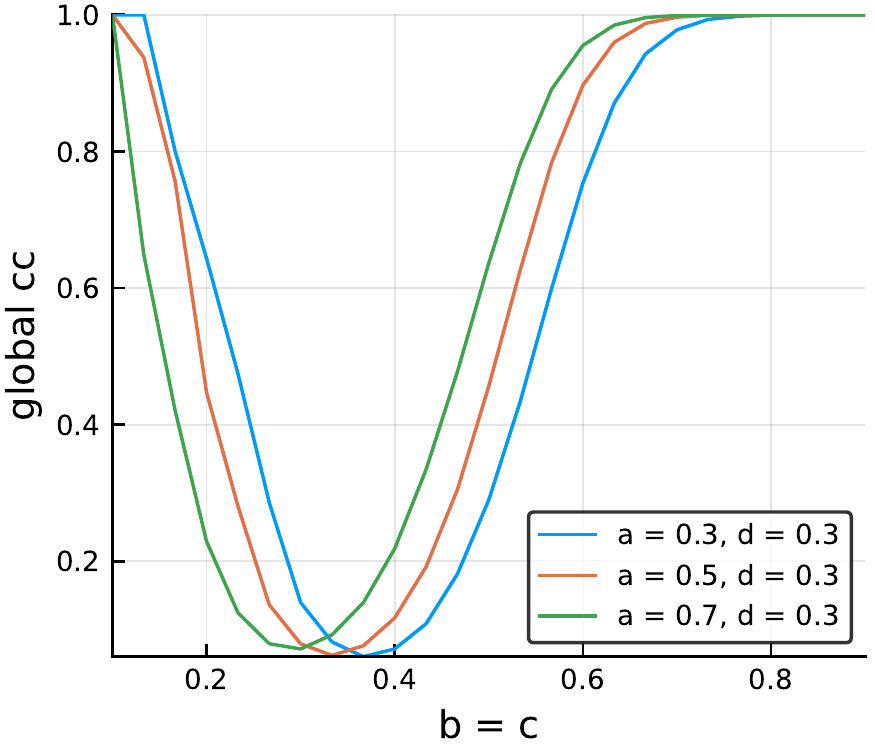}
	%\vspace*{-2.0\baselineskip}
	\caption{\footnotesize Global clustering coefficients vary with changing HyperKron Parameters. Here $r = 10$.}
	%\vspace*{-\baselineskip}
	\label{fig:clust}
\end{marginfigure}

Next we present the case the HyperKron model allows for generating models with significant \emph{clustering} even with few edges. This is an improvement over the original Kronecker model, which will be explored further in \S\ref{sec:fitting}. We use the global clustering coefficient: $ 6|K_3| / |W| $, where $|K_3|$ is the number of triangles, and $|W|$ is the number of wedges, to measure how much network nodes tend to form triangles~\cite{watts1998collective}. 
We generate the HyperKron model for fixed $a$ and $d$ parameters, using $r=10$. Figure \ref{fig:clust} demonstrates that for varying all values of $b$ and $c$ the global clustering is always above 0.05, and is often much larger. It is large initially because all edges are in triangles. As the network becomes denser ($b,c$ get larger), the wedges emerge causing the coefficient to drop.  Finally, as the network becomes quite dense, these wedges combine into triangles. But throughout, clustering remains. This is significant because we can still achieve good clustering with sparse networks (the real-world behavior) with the HyperKron model.

\section{Fitting HyperKron to Real Data}
\label{sec:fitting}
We demonstrate now that the HyperKron model can be fit to real-world data by hand-tuning the coefficients. Four real-world networks were chosen: \textit{email} comes from Arena's collection, and is a list of email exchanges between members of the University Rovira i Virgili with 1133 nodes~\cite{Guimera-2003-interactions}; \textit{Villanova62} (7772 nodes) and \textit{MU68} (15k nodes) come from the facebook 100 data set~\cite{Traud-2012-facebook} where nodes represent people and edges are friendships; and \textit{homo} is a biology network of protein interactions with 8887 nodes~\cite{singh2008-isorank-multi}.

To fit real-world data to our HyperKron model, we choose to fit the model to just the set of triangles in the network as this is the natural structure for HyperKron to generate. (See \S\ref{sec:degree} where we consider the full network.) Fitting the coefficients to a symmetric HyperKron model with a $2\times 2\times 2$ initiator matrix (four parameters) was done by hand. See our choices $a,b,c,$ and $d$ in Table~\ref{tab:Fit}.

For comparison, we fit the same data sets to the Kronecker model as well using the method-of-moments~\cite{Gleich-2012-kronecker} (called KGMome in the table) and maximum likelihood~\cite{leskovec2010_Kronecker} (called KGFit in the table). (These methods often picked different $r$.) For reference, we also fit those models to the full edge data in addition to the extracted triangle data.  While there are other models that would also capture clustering~\cite{Kolda-2014-BTER,newman2009random}, these require far more parameters and so we don't compare against them.

\subsection{Clustering Coefficients}
\label{sec:clust}
The traditional global clustering coefficient is described in \S\ref{sec:non-trivial}. The average local clustering coefficient is the average over the local clustering coefficient defined for each node u: $ 2 |K_3(u)| / |W(u)|,$ where $|K_3(u)|$ is the number of triangles for which $u$ is a member, and $|W(u)|$ is the number of wedges for which $u$ is a member.
One of the biggest improvements of the HyperKron model over other graph models such as the original Kronecker model, is the ability to capture clustering in the model. Regardless of using the full data, or restricted triangle data, the Kronecker models do not capture clustering properties as closely HyperKron model does (see Table~\ref{tab:Fit}).

%Although the HyperKron model captures clustering, it the lack of nodes outside of triangles.
There remain properties of the real-world networks that HyperKron does not possess. For instance, the HyperKron model lacks higher-order clustering. We use the methodology and code presented in~\citet{yin2017higher} to compute \textit{higher order clustering coefficients}. The precise details are not relevant for our case, but these extend clustering coefficients to larger cliques. 
%The $l$-th order global clustering coefficient is $ (l^2 +l) |K_{l+1} | / |W_l|$, where $|K_{l+1}|$ is the number of $(l+1)$-cliques, and $|W_l|$ is the set of $l$-wedges. (An $l$-wedge is a $l-1$ clique plus one edge). Similarly, the local clustering coefficient is the average over the local clustering coefficients for each node: $ |K_{l+1}| / |W_{l}(u)|$. 
%Note that our earlier definition of clustering corresponds with the $2$nd order clustering coefficient. 
We find that the HyperKron model does not display clustering in terms of four cliques, five cliques, or six cliques (3rd, 4th, and 5th order) and the coefficients for higher orders are small, as seen in Table~\ref{tab:higherclust}.

% KG_Mome = David's fitting
% KG_Fit = Jure's fitting
\newcolumntype{L}[1]{>{\hsize=#1\hsize\raggedright\arraybackslash}X}%
\newcolumntype{R}[1]{>{\hsize=#1\hsize\raggedleft\arraybackslash}X}%
\newcolumntype{C}[1]{>{\hsize=#1\hsize\centering\arraybackslash}X}
\begin{tuftetable}
	\caption{Fitting real world data to the HyperKron model. Note that the only model with non-trivial global and local clustering are the HyperKron (HKron) fits. See the text for some of the details.
		We list the number of nodes, number of edges, global clustering coefficient, mean-local clustering coefficient, and the size of the largest connected component.}
	%\vspace*{-\baselineskip}
	\small
	\begin{tabularx}{\columnwidth}{ L{3} L{.5} L{.5} L{.5} L{.5}}
		\toprule
		&           & global & local & lcc \\
		Network name       & edges &  clust & clust & size \\ \midrule
		\textbf{\textit{email} full} & 5451 & 0.166 & 0.220 & 1133 \\ 
		{\footnotesize KGFit: $(.9538,.6196,.1463)$ r = 11} & 4941  & 0.032 & 0.060 & 1803 \\
		{\footnotesize KGMome: $(1.0 , 0.5241 , 0.2990)$, r = 11} & 5945 & 0.035 & 0.031 & 1351 \\ \addlinespace
		
		\textbf{\textit{email} triangles}  & 4229 & 0.232 & 0.366 & 837\\
		\footnotesize{HKron: $(0.999,0.31, 0.2,0.0001)$, r = 10}  & 4546 & 0.140 & 0.346 & 735\\
		\footnotesize{KGFit: $(.9036,.6946,.2056)$, r = 10}  & 4736&  0.052 & 0.076 & 949 \\ 
		\footnotesize{KGMome: $(1.0 , 0.5132 , 0.2688 )$, r = 11} & 4651& 0.034 & 0.032 & 1393 \\ \midrule
		
		\textbf{\textit{homo}  full} & 33k & 0.070 & 0.133 & 8887 \\
		\footnotesize{KGFit: $(.9895,.5569,0.1147)$, r = 14} &34k & 0.013 & 0.025 & 6547 \\
		\footnotesize{KGMome: $(1.0 , 0.5676 , 0.0759)$, r=14} & 33k & 0.015 & 0.033 & 6333 \\ \addlinespace
		
		\textbf{\textit{homo} triangles}  & 19k & 0.141 & 0.264 & 3783\\
		\footnotesize{HKron: $(0.8,0.115, 0.15,0.83)$, r = 12} & 19k & 0.101 & 0.164 & 4072\\
		\footnotesize{KGFit: $(.9487,.6416,.1832  )$, r = 12} & 20k & 0.027  & 0.048  & 3194\\ 
		\footnotesize{KGMome: $( 1.0 , 0.5227 , 0.0882 )$, r=14} & 20k & 0.013 & 0.022 & 4502  \\ \midrule
		
		\textbf{\textit{Villanova62} full} & 315k & 0.166  & 0.235& 7755\\
		\footnotesize{KGFit: $(.9999,.7064,.388)$, r = 13} & 326k & 0.056 & 0.064 & 8187 \\
		\footnotesize{KGMome: $( 1.0 , 0.696 , 0.4086)$, r = 13} & 326k & 0.054 & 0.059 & 8185 \\ \addlinespace
		 
		\textbf{\textit{Villanova62} triangles}  & 311k & 0.168  & 0.258 & 7476 \\
		\footnotesize{HKron: $(0.9,0.4,.24,.001)$, r = 13} & 306k & 0.111& 0.265 & 7944   \\
		\footnotesize{KGFit: $( 0.9999, .7058, .3865 )$, r = 13} & 322k& 0.055 & 0.064 & 8187 \\ 
		\footnotesize{KGMome: $( 1.0 , 0.6965 , 0.4054  )$, r = 13} & 323k &0.054 & 0.059 & 8186\\ \midrule
		
		\textbf{\textit{MU78} full} & 649k & 0.152 & 0.214 & 15k \\
		\footnotesize{KGFit: $(.996,.675,.3992)$, r = 14} & 690k & 0.034 & 0.037 & 16k  \\ 
		\footnotesize{KGMome: $( 1.0 , 0.6305 , 0.4790  )$, r = 14} & 672k & 0.028 & 0.026 & 16k \\ \addlinespace
		
		\textbf{\textit{MU78} triangles}  & 637k & 0.155 & 0.240 & 15k\\
		\footnotesize{HKron: $(0.9,0.42,0.20,0.001)$, r = 14}  & 625k & 0.097 & 0.295 & 16k \\
		\footnotesize{KGFit: $( 0.9993, 0.6721, 0.3973 )$, r = 14} & 675k & 0.037 & 0.034 & 16k\\ 
		\footnotesize{KGMome: $( 1.0 , 0.6311 , 0.4745 )$, r = 14} & 661k &0.028 & 0.026 & 16k\\ \bottomrule
	\end{tabularx}
	\label{tab:Fit}
\end{tuftetable}

\begin{tuftetable}
	\caption{Higher order global clustering coefficients.}
	%\vspace*{-\baselineskip}
		\small
	\begin{tabularx}{\linewidth}{ L{2.95} L{.35} L{.35} L{.35}}
		\toprule
		Network name       & $3^{rd}$&  $4^{th}$ & $5^{th}$ \\ \midrule
		\textbf{\textit{email} triangles} & 0.137 & 0.156  & 0.223\\
		\footnotesize{HKron: $(0.999,0.31, 0.2,0.0001)$, r = 10}  & 0.065 & 0.045 & 0.033 \\
		\textbf{\textit{homo} triangles} & 0.113 & 0.184 & 0.261  \\
		\footnotesize{HKron: $(0.8,0.115, 0.15,0.83)$, r = 12} & 0.002 & 0.0 & 0.0 \\
		\textbf{\textit{Villanova62} triangles} &  0.109 & 0.115 & 0.131 \\
		\footnotesize{HKron: $(0.9,0.4,.24,.001)$, r = 13} & 0.050 & 0.037 & 0.031 \\
		\textbf{\textit{MU78} triangles} & 0.137 & 0.164 & 0.175 \\
		\footnotesize{HKron: $(0.9,0.42,0.20,0.001)$, r = 14} & 0.052 & 0.040 & 0.033 \\
		\bottomrule
	\end{tabularx}
	\label{tab:higherclust}
\end{tuftetable}

\normalsize
\subsection{Skewed Degrees}
\label{sec:degree}
Another desirable property that the HyperKron model preserves is a highly skewed degree distribution. That is, there are a few nodes with very large degree, with the average degree being much smaller. Figure~\ref{fig:degree} shows the degree distributions in log-scale for two of our networks: \textit{Villanova62}, and \textit{MU78}, along with their HyperKron fits. We also show Loess smoothed estimates to show broader properties.  

There are two notable behaviors in the HyperKron degree distribution. First, there are two ``tails'' in nodes with lower degree. The tail with larger counts are nodes with even degree. They occur in higher frequency since the model most often adds two neighbors to a node at once when we add triangles. Conversely, the tail with smaller counts are nodes with odd degree, since a single edge is placed infrequently.  Second, the HyperKron model shows an interesting pattern in the high-degree vertices. This is a known problem with Kronecker models and occurs in the original version as well. The peaks can be smoothed out in the Kronecker model by perturbing the probability matrix as demonstrated in~\citet{Seshadhri-2013-kronecker}. We implemented a similar procedure which we explain next.

\begin{tuftefigure}
\centering
		\includegraphics[width = 0.4\textwidth]{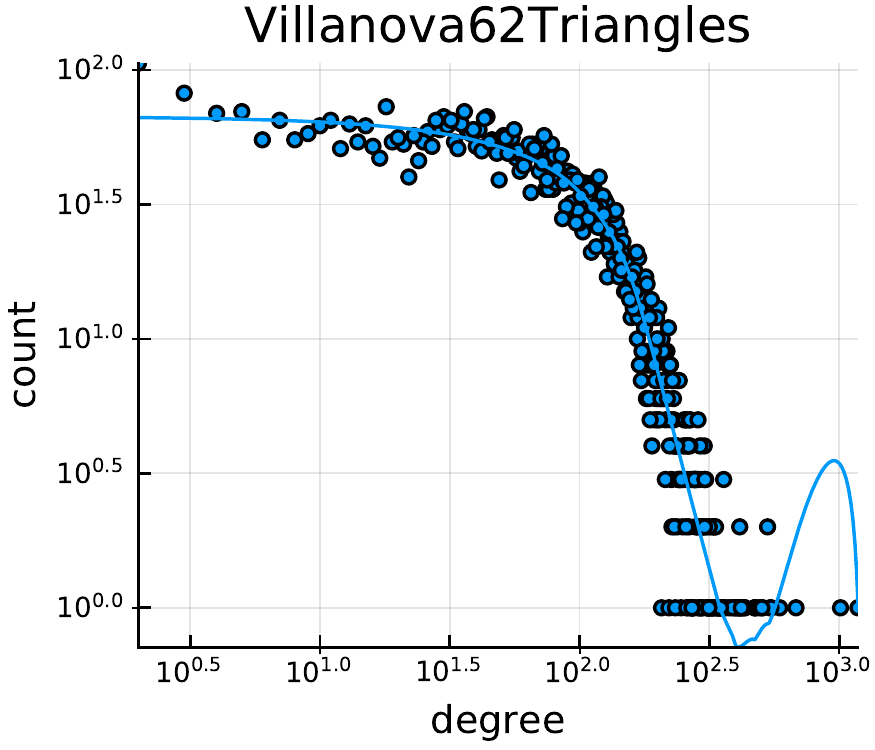}
		\includegraphics[width = 0.4\textwidth]{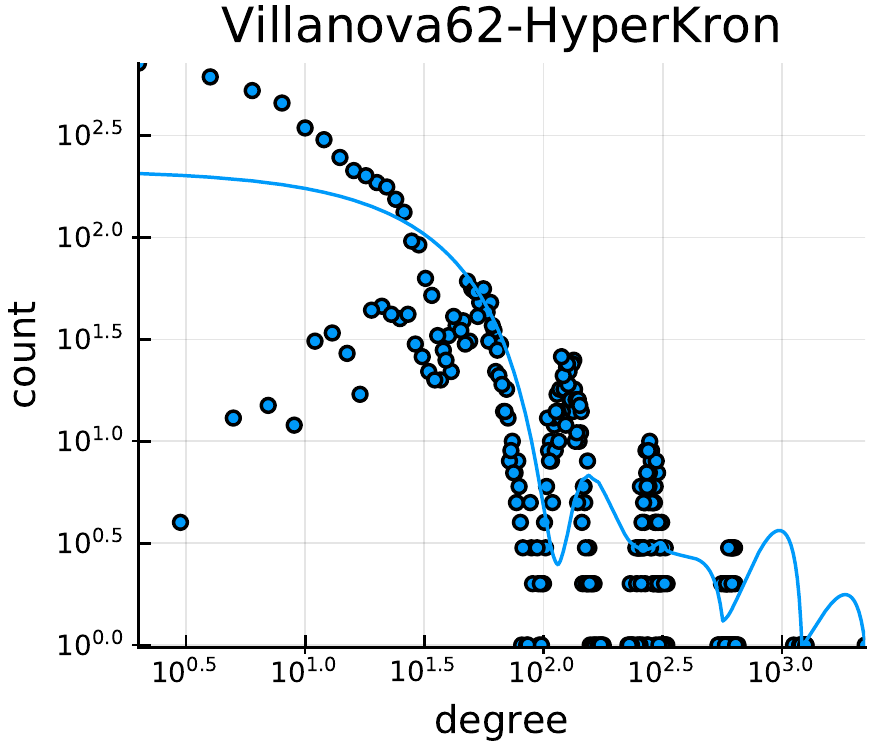}
		\includegraphics[width = 0.4\textwidth]{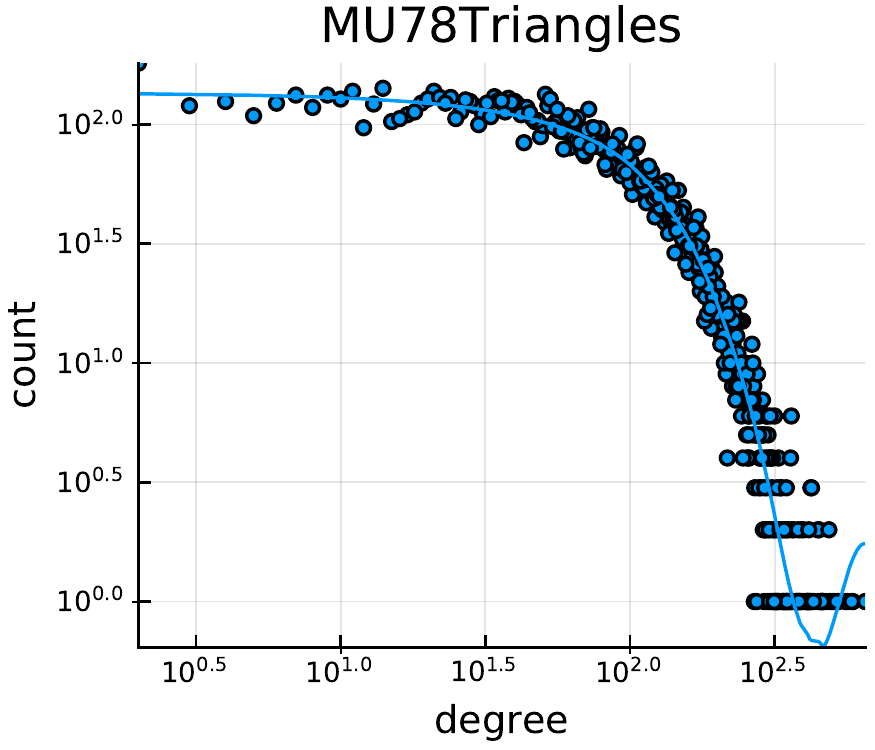}
		\includegraphics[width = 0.4\textwidth]{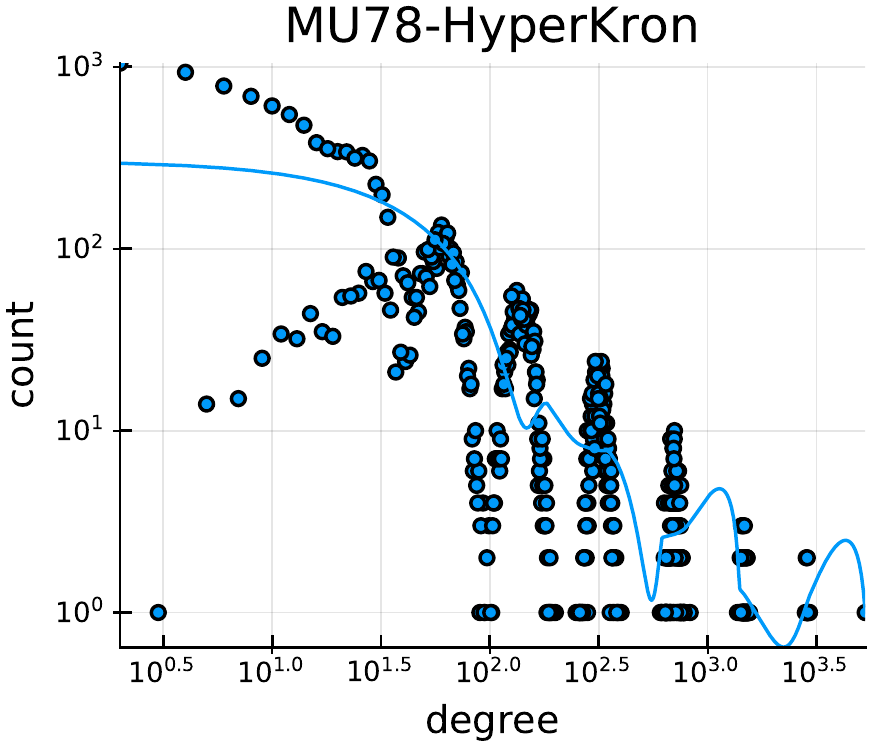}
\vspace*{-\baselineskip}
	\caption{HyperKron preserves highly skewed degree distribution with notable behaviors discussed in the text.}
	\label{fig:degree}
\end{tuftefigure}

We made several tweaks to the HyperKron generation to address these issues. First of all, we add a \emph{noise} parameter to the HyperKron model in a generalization of the method in~\citet{Seshadhri-2013-kronecker}. Recall that the probability of placing a particular hyperedge is the multiplication of $r$ entries of the initiator tensor, $\cmP$. To incorporate noise we perturb each of the $r$ tensors involved in the Kronecker product with two noise parameters, $\mu_i, \nu_i$, in the following way. Instead of using the tensor in \eqref{eq:sym-tensor-P}, for each level $i = 1,2, \ldots r$ we use
\begin{smalleq}%
\begin{equation}
\cmP_i =
\left[
\begin{array}{cc|cc}
a - \frac{3a \mu_i}{a+d} - \frac{3a \nu_i}{a+d} & b + \mu_i & b + \mu_i & c + \nu_i \\
b+ \mu_i & c + \nu_i & c + \nu_i & d  - \frac{3d \mu_i}{a+d} - \frac{3d \nu_i}{a+d}
\end{array}
\right],
\end{equation}
\end{smalleq}%
where $\mu_i, \nu_i$ are uniformly randomly sampled within an appropriate range, $[-\sigma,\sigma]$ and $\sigma \le \min(b,c)$ (akin to the case for Kronecker in~\citet{Seshadhri-2013-kronecker}). Using this added noise parameter, we fit HyperKron to the set of edges involved in triangles, using the same initiator parameters as before. 

The second adjustment that is to also account for the set of remaining edges (those not involved in triangles).  We fit this residual set of edges to a Kronecker model using the method of moments in~\citet{Gleich-2012-kronecker}, with an added noise parameter as in~\citet{Seshadhri-2013-kronecker}. Note that when we add the Kronecker graph to the HyperKron graph, many of the edges overlap. So finally, we add in an \ERfull graph with an expected number of edges set to add enough edges to get back to the number of edges of the original graph. Adding on additional edges in these ways is natural, as both the Kronecker and \ERfull graphs can be derived from subsets of the HyperKron tensor where two indices are equal. (That is, we can view these as instances of sums of HyperKron models, but we don't pursue that formalism here.) 

\begin{tuftefigure}
	\centering
	\includegraphics[width = 0.4\textwidth]{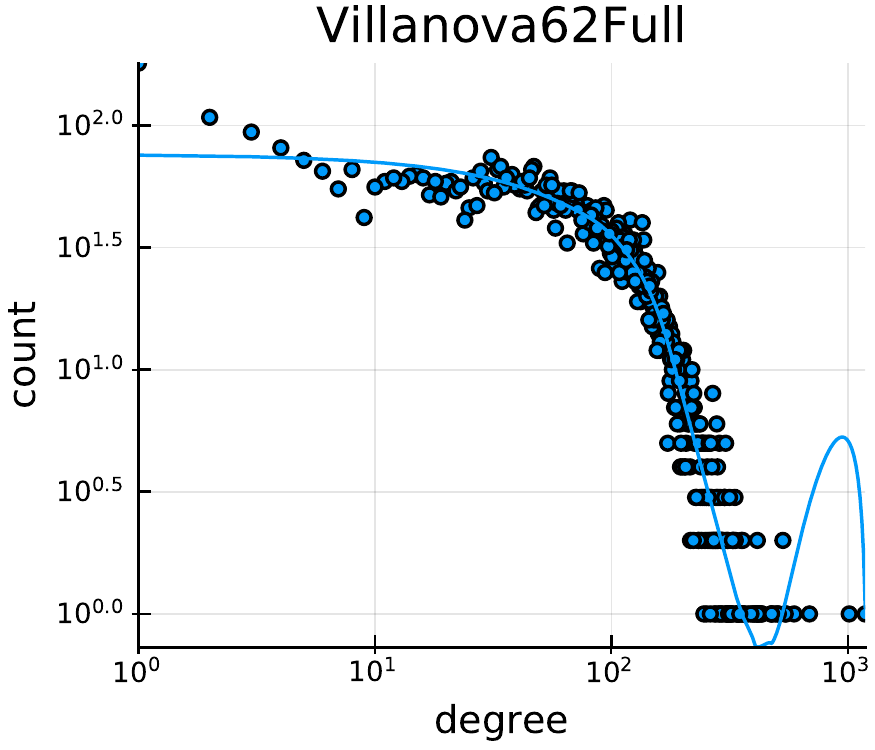}
	\includegraphics[width = 0.4\textwidth]{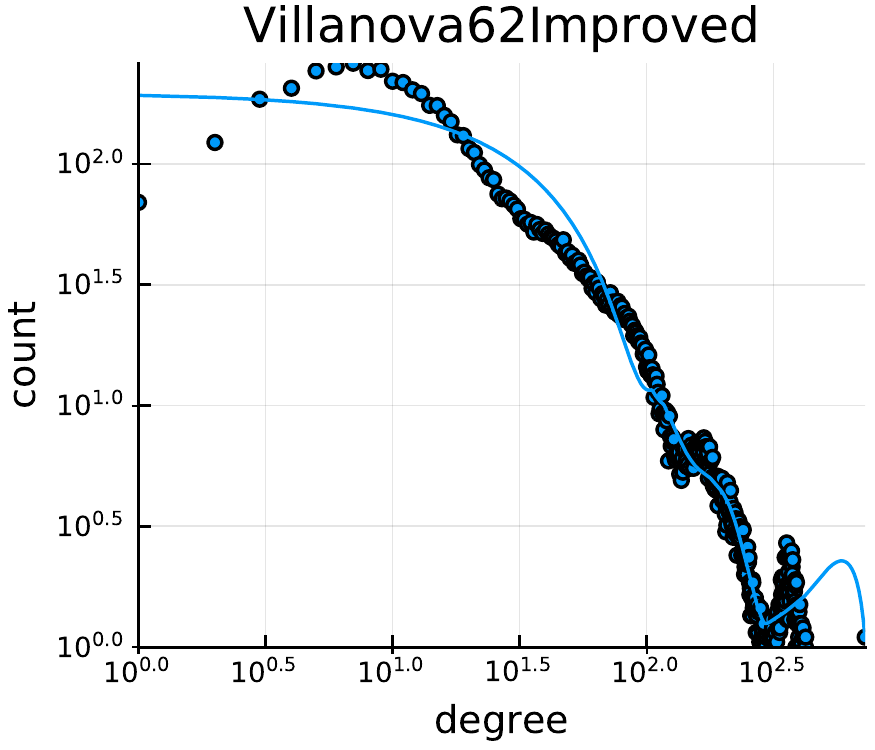}
	\includegraphics[width = 0.4\textwidth]{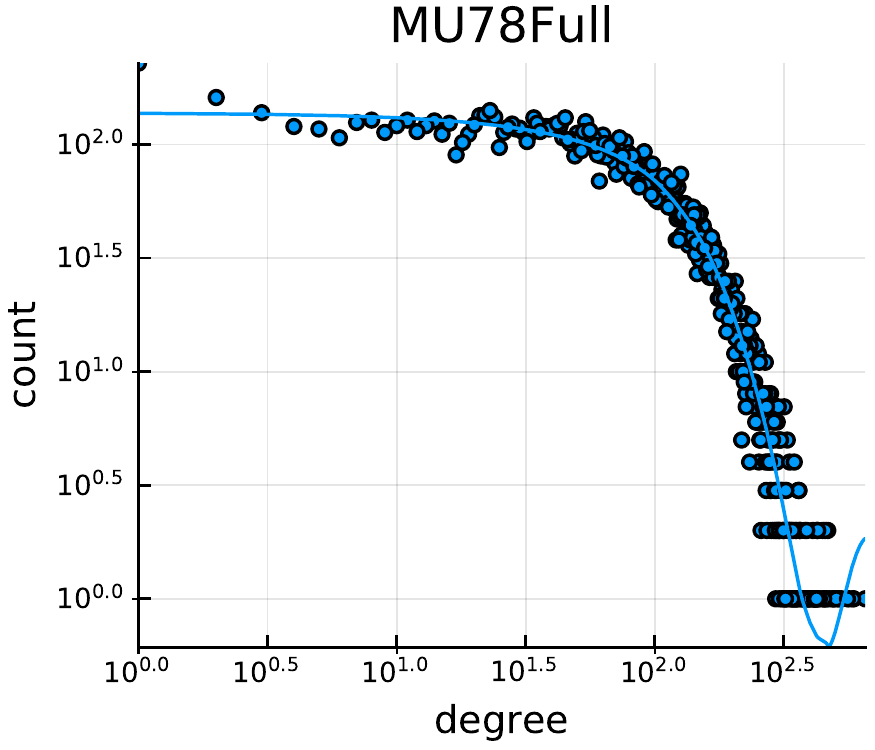}
	\includegraphics[width = 0.4\textwidth]{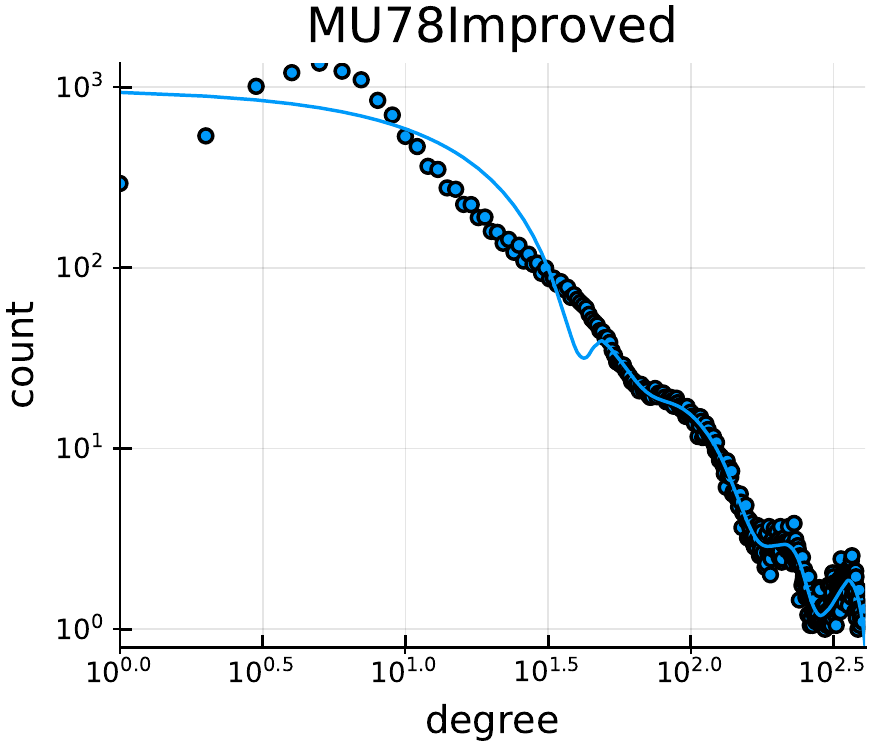}
	%\vspace*{-\baselineskip}
	\caption{Improvements to the HyperKron model eliminates two-tailed behavior and almost entirely removes oscillation and improves the fit to data. See the text for details.}
	\label{fig:noise}
\end{tuftefigure}

Figure~\ref{fig:noise} gives the degree distributions of the full original network data in log-scale, along with the improved fitting. For Villanova62, the HyperKron noise ($\sigma$) was set to 0.15, and the Kronecker noise was set to 0.1. For MU78 the HyperKron noise was set to .2 and the Kronecker noise was set to 0.05. The two tailed behavior is eliminated by fitting to the non-triangle edges, and the oscillation behavior is almost entirely removed by the noise. In both cases, the fittings retain non-trivial clustering coefficients.
\section{Model flexibility}
\label{sec:flex}

Thus far, HyperKron was described in a setting where triangles are associated with each generated hyperedge. As we have seen, this is an appropriate choice for settings where we expect 2nd order (triangle-based) clustering in undirected networks. There are more complex types of network data, and
we now show that the HyperKron model is also relevant for these more interesting data. 

\begin{marginfigure}
\includegraphics[width=\linewidth]{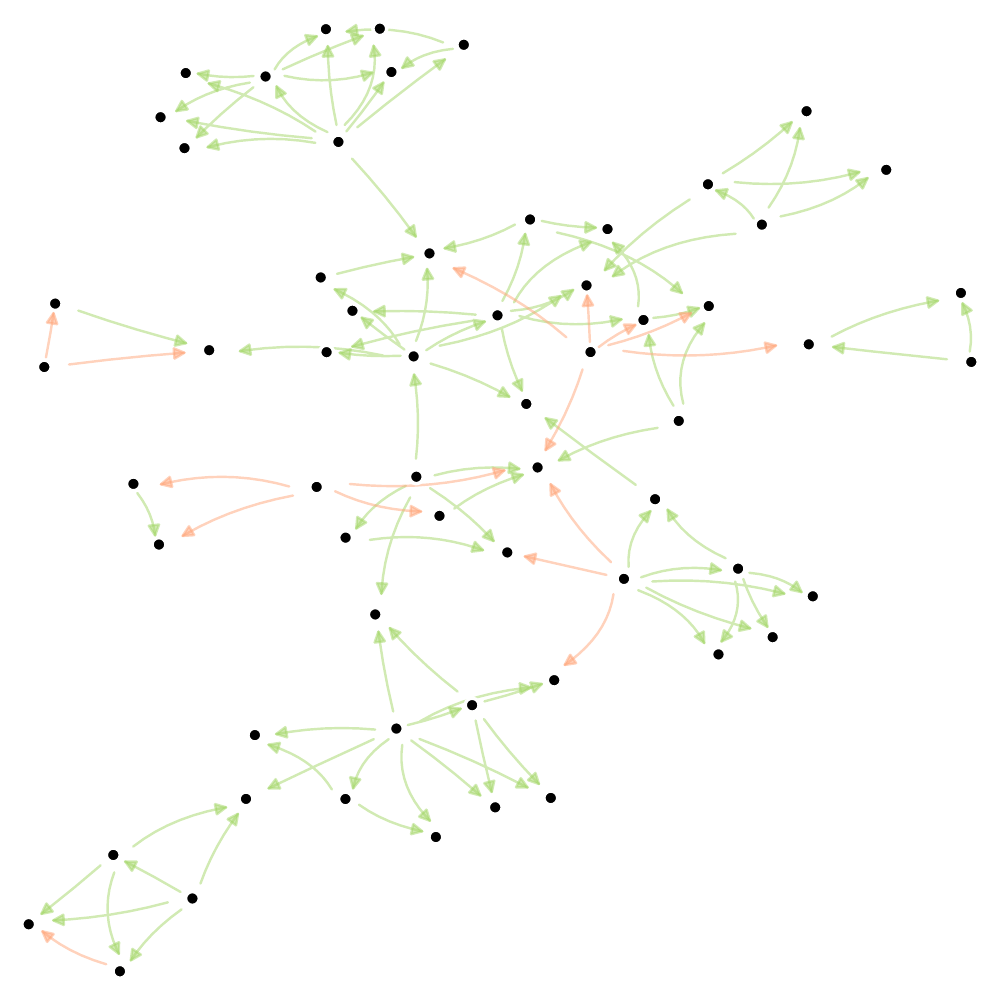}
\includegraphics[width=\linewidth]{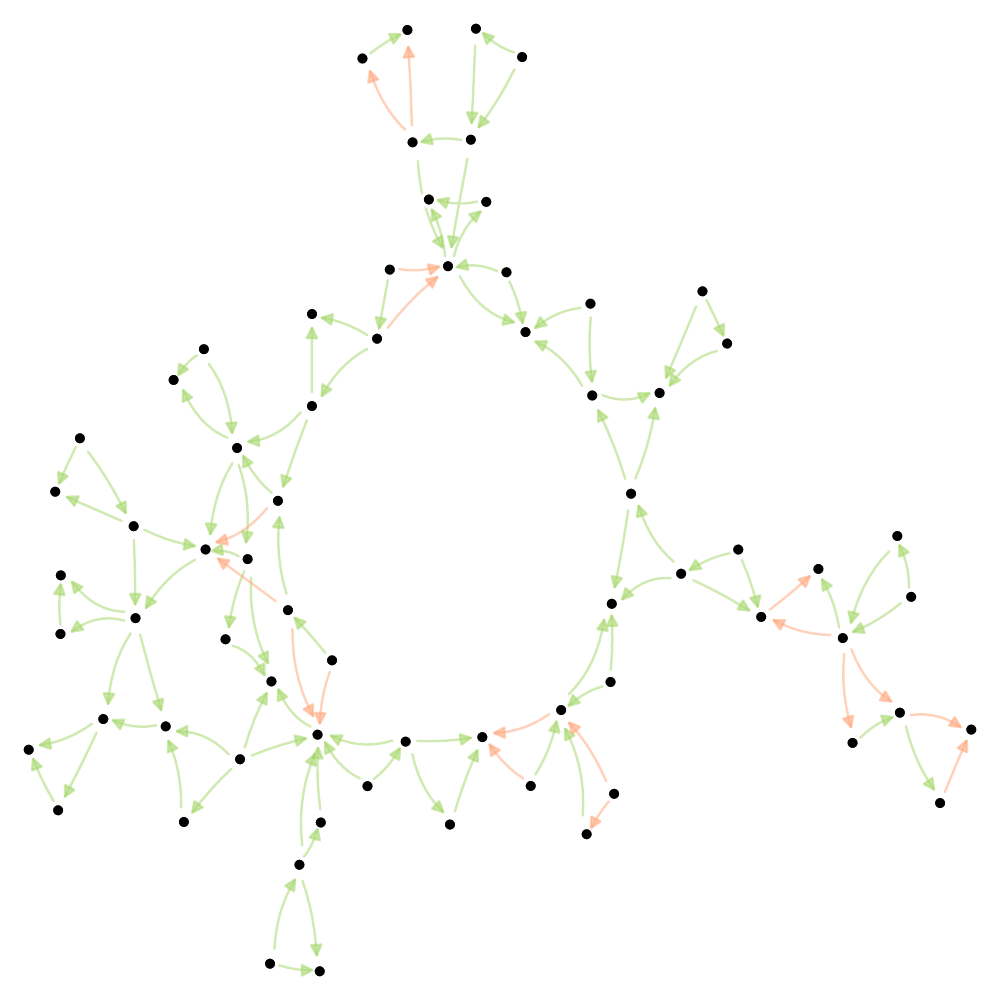}
%\vspace*{-1.5\baselineskip}
\caption{At top, a network drawing of the nodes involved in the feed forward
loops in \emph{S.~cerevisiae}. Green edges denote promotion (positive sign) 
and orange edges denote repression (negative sign). At bottom, a network drawing 
of the HyperKron model described in the text.}
\label{fig:yeast}
\end{marginfigure}

For instance, the \emph{S.~cerevisiae} transcription regulatory network 
is a directed, signed graph that describes promotion or repression of 
gene expression in the common yeast organism. Coherent feed-forward loops
are an important higher-order structure in this network~\cite{milo_network_motifs}.
We extract all nodes involved
in coherent feed forward loops, leaving a network with 61 nodes and 108 
directed, signed edges (92 positive, 16 negative). By manually tweaking
entries to get the number of edges to match, we generated HyperKron model using a $2 \times 2 \times 2$ tensor
with parameters: 
%\[
%\begin{array}{@{}>{\footnotesize}l>{\footnotesize}l>{\footnotesize}l>{\footnotesize}l@{}}
%(1,1,1) \leftarrow 0.14 & (1,2,1) \leftarrow 0.25 & (2,1,1) \leftarrow 0    & (2,2,1) \leftarrow 0.45 \\
%(1,1,2) \leftarrow 0.55 & (1,2,2) \leftarrow 0    & (2,1,2) \leftarrow 0.31 & (2,2,2) \leftarrow 0.06
%\end{array}
%\] 
\begin{smalleq}%
	\begin{equation}%
	\begin{array}{llll}%
	\cmP_{111} = 0.14 &  \cmP_{121} = 0.25& \cmP_{211} =0 & \cmP_{221} = 0.45 \\
	\cmP_{112} = 0.55 & \cmP_{122} = 0 & \cmP_{212} = 0.31 &  \cmP_{222} = 0.06 \nonumber
	\end{array}%
	\end{equation}%
\end{smalleq}% 
and $r = 7$. 
We associated each hyperedge with one of the four coherent feed-forward loops
based on a biased random choice. More specifically, the type 1,
all positive motif had probability 1/2, the type 2 motif had probability 1/4,
the type 3 and 4 motifs had probability 1/8, respectively (see~\citet{milo_network_motifs}
for more about these types). These were chosen
because the real network doesn't have any type 3 and 4 feed-forward loops. 
When we assemble the motifs placed via these hyperedges into a network, 
any two motifs that share an edge with the same direction will be 
coalesced by summing the signs. 
The largest connected component of the resulting network had 69 nodes and 108
directed, signed edges (90 positive, 18 negative). 

We show graph drawings of the two networks in Figure~\ref{fig:yeast}
The real network has 38 coherent feed forward loops and 2 incoherent feed
forward loops. The HyperKron model has 36 coherent feed forward loops and
1 incoherent feed forward loop. Note that the presence of incoherent feed
forward loops is an emergent behavior because we only ever generated coherent
loops. In this case, we might ask if finding 2 incoherent feed forward loops
in the real network is likely to occur or not. By generating 10000 instances
of our model, we find at least 2 incoherent feed forward loops around 1006
times (roughly 10\%). Consequently, the presence of these two loops in the
real data could easily have occurred by chance.  Our code to reproduce
this experiment will be posted online.

\section{Related \& future work \& discussion}
\label{sec:related}

Let us be clear that we do not believe that HyperKron is a universal network model that is always appropriate. Rather, it provides some complementary directions in the large space of network models. 
%\footnote{Note that we are unable to provide references for all the models in this section due to space restrictions. We add citations to models that are not well-known.} 
One of the advantages of hypergraph based modeling such as HyperKron is that it provides an easy and flexible means to incorporate higher-order structure. This was used as well in \citet{Bollobas-2011-hypergraphs} where they associated hyperedges with triangles.  We use this flexibility to model directed, signed networks in \S\ref{sec:flex} and networks with non-trivial clustering coefficients in \S\ref{sec:clust}. It is not obvious how to generate these types of structures for models based on matrices of probabilities such as \ERfull, Chung-Lu, or kernel functions~\cite{Hagberg2015}. The same critique holds for evolutionary models such as the copying model or forest-fire model. And in addition, the HyperKron model is easy to simulate in parallel -- you can parallelize over the \ERfull regions, for instance.

That said, there are other types of network models that possess clustering. \citet{newman2009random} studied a configuration model that incorporated the triangle degree of each node. \citet{Kolda-2014-BTER} proposed the BTER model that has large clustering coefficients and a reasonable match. These are both excellent models with clustering, but is unclear how to incorporate more complex types of structure such as signs into these models. Likewise, models that randomly generate points for each node and then connect nearby nodes based on a metric space are often known to have non-trivial local clustering~\cite{Bonato-2012-geop,jacob2015spatial}. However, these models tend to be unrealistically dense if the geometry is not sufficiently high dimensional, at which point you lose local clustering. HyperKron is related to another generalization of the Kronecker Model, RTM~\cite{akoglu2008rtm}, in the sense that it too uses a 3 dimensional tensor, but RTM does not incorporate higher-order structure like HyperKron.

Our work also continues to evolve the space of Kronecker models. In fact, our HyperKron model is fairly easy to combine with the majority of other ideas that have been proposed to extend Kronecker models. It would be easy, for instance, to adapt the mKPGM model~\cite{Moreno-2010-mKPMG} to our setting as it simply involves a deterministic choice for some of the early tensors. Likewise, the MAG model uses a set of Kronecker models to handle attributed graphs~\cite{kim2010multiplicative}.

%As we have indicated throughout, there are meaningful deviations between the behavior of HyperKron models and real data. This is shown most vividly in the degree distribution plot (Figure~\ref{fig:degree}). Seshadhri et al.~\cite{Seshadhri-2013-kronecker} proposed a way to address this type of bias in the degree distribution with a ball-dropping perspective on Kronecker graphs by adding a particular type of noise. We were unable to adapt their procedure to our setting, and so this is one of our most important next steps. 

%Some of the other issues with the model that remain open include how to combine HyperKron models with models that will place more individual edges in the graph. This would allow us to avoid fitting the HyperKron model to only regions that are rich in higher-order structure. It turns out that by restricting the tensor $\cmP^r$ to only those entries that would generate an edge reveals an embedded Kronecker graph. We believe that this feature can be used to \emph{add} a Kronecker graph models of individual edges to generate more interesting graphs. Once this is done, it will be more difficult to hand-tune a model fit to real data and we will need to develop automated fitting techniques. 

One issue with this model that remains open includes how to pragmatically fit HyperKron to real data. We were able to hand tune parameters in \S\ref{sec:fitting}, by getting the number of edges to match. However it would be nice to have an automated fitting technique similar to~\citet{Gleich-2012-kronecker} or~\citet{leskovec2010_Kronecker}. As discussed in \S\ref{sec:analytic}, there are a number of duplicate edges placed in the HyperKron model, which makes it difficult to estimate features of HyperKron, so this area remains one of our most important next steps.

We also illustrated the lack of higher-order clustering in the HyperKron models. We believe this aspect is a feature of the model as it enables testing specific sources of clustering structure. For instance, if the goal is to test hypotheses about 2nd order clustering structure in the network, then the lack of higher-order structure is useful. Additionally, the HyperKron model would also extend to larger size motifs such a four-clique and five-cliques through a suitable adaptation of the HyperKron sampling procedure to 4th and 5th order tensors. These models will exhibit higher-order clustering. We also plan to explore these settings in the future.

Let us conclude by reiterating that the HyperKron model has a number of useful features in the space of network models. We have illustrated many of these throughout our manuscript. Like the Kronecker model, it is a simple, low-parameter, and flexible model that is relatively easy to analyze---as we have done for a few features. It scales up to large networks and can model some aspects of real-world networks.

%\balance
%\bibliographystyle{abbrv}
\begin{fullwidth}
\bibliographystyle{dgleich-bib}
\bibliography{99-refs,All_Docs}
\end{fullwidth}
%\bibliography{99-refs}

%This must be removed for ICDM submission
%\footnotesize
%\noindent \textbf{Acknowledgements.} 
%	The authors were supported by NSF CCF-1149756, IIS-1422918, IIS-1546488, CCF-0939370, DOE DE-SC0014543, DARPA SIMPLEX, and the Sloan Foundation.

\end{document}